\documentclass[fleqn,usenatbib]{mnras}

\usepackage{newtxtext,newtxmath}

\usepackage[T1]{fontenc}

\DeclareRobustCommand{\VAN}[3]{#2}
\let\VANthebibliography\thebibliography
\def\thebibliography{\DeclareRobustCommand{\VAN}[3]{##3}\VANthebibliography}

\usepackage{graphicx}	
\usepackage{amsmath}	

\usepackage{hyperref}
\hypersetup{
    colorlinks=true,
    linkcolor=blue,
    citecolor=blue,
    filecolor=magenta,      
    urlcolor=cyan,
    }
\usepackage{textgreek}
\usepackage{pdflscape}
\usepackage{booktabs}
\usepackage{amsmath}
\usepackage{caption}
\usepackage{placeins}
\usepackage{capt-of}
\usepackage{csvsimple}

\newcommand{\lya}{Ly\textalpha}
\newcommand{\ha}{H\textalpha}
\newcommand{\hb}{H\textbeta}
\newcommand{\hg}{H\textgamma}
\newcommand{\oiii}{[O\,\textsc{iii}]}
\newcommand{\oii}{[O\,\textsc{ii}]}
\newcommand{\hii}{H\,\textsc{ii}}
\newcommand{\ciii}{C\,\textsc{iii}]}
\newcommand{\oiiiuv}{O\,\textsc{iii}]}
\newcommand{\civ}{C\,\textsc{iv}}
\newcommand{\heii}{He\,\textsc{ii}}

\newcommand{\xiion}{$\xi_\mathrm{ion}$}

\newcommand{\fesc}{$f_\mathrm{esc}$}
\newcommand{\muv}{$M_\mathrm{UV}$}

\newcommand{\flux}{erg\,s$^{-1}$\,cm$^{-2}$}

\title[Spectroscopic UV continuum slopes at high-$z$]{Hitting the slopes: A spectroscopic view of UV continuum slopes of galaxies reveals a reddening at $\mathbf{z>9.5}$}

\author[A. Saxena et al.]{Aayush Saxena,$^{1,2}$\thanks{E-mail: aayush.saxena@physics.ox.ac.uk}
Alex J. Cameron,$^{1}$
Harley Katz,$^{3,4}$
Andrew J. Bunker,$^{1}$
Jacopo Chevallard,$^{1}$
\newauthor 
Francesco D'Eugenio,$^{5,6}$
Santiago Arribas,$^{7}$
Rachana Bhatawdekar,$^{8}$
Kristan Boyett,$^{1}$
Phillip A. Cargile,$^{9}$
\newauthor 
Stefano Carniani,$^{10}$
St\'{e}phane Charlot,$^{11}$
Mirko Curti,$^{12}$
Emma Curtis-Lake,$^{13}$
Kevin Hainline,$^{14}$
\newauthor 
Zhiyuan Ji,$^{14}$
Benjamin D. Johnson,$^{9}$
Gareth C. Jones,$^{5,6}$
Nimisha Kumari,$^{15}$
Isaac Laseter,$^{16}$
\newauthor 
Michael V. Maseda,$^{16}$
Brant Robertson,$^{17}$
Charlotte Simmonds,$^{5,6}$
Sandro Tacchella,$^{5,6}$
Hannah \"Ubler,$^{18}$
\newauthor 
Christina C. Williams,$^{19}$
Chris Willott,$^{20}$
Joris Witstok,$^{21,22}$
Yongda Zhu$^{14}$
\\
$^{1}$Department of Physics, University of Oxford, Denys Wilkinson Building, Keble Road, Oxford OX1 3RH, UK \\
$^{2}$Department of Physics and Astronomy, University College London, Gower Street, London WC1E 6BT, UK \\
$^{3}$Department of Astronomy \& Astrophysics, University of Chicago, 5640 S Ellis Avenue, Chicago, IL 60637, USA\\
$^{4}$Kavli Institute for Cosmological Physics, University of Chicago, Chicago IL 60637, USA\\
$^{5}$Kavli Institute for Cosmology, University of Cambridge, Madingley Road, Cambridge CB3 0HA, UK \\
${^6}$Cavendish Laboratory, University of Cambridge, 19 JJ Thomson Avenue, Cambridge CB3 0HE, UK\\
$^{7}$Centro de Astrobiolog\'ia (CAB), CSIC–INTA, Cra. de Ajalvir Km.~4, 28850- Torrej\'on de Ardoz, Madrid, Spain\\
$^{8}$European Space Agency (ESA), European Space Astronomy Centre (ESAC), Camino Bajo del Castillo s/n, 28692 Villanueva de la Cañada, Madrid, Spain\\
$^{9}$Center for Astrophysics $|$ Harvard \& Smithsonian, 60 Garden St., Cambridge MA 02138 USA\\
$^{10}$Scuola Normale Superiore, Piazza dei Cavalieri 7, I-56126 Pisa, Italy\\
$^{11}$Sorbonne Universit\'e, CNRS, UMR 7095, Institut d'Astrophysique de Paris, 98 bis bd Arago, 75014 Paris, France\\
$^{12}$European Southern Observatory, Karl-Schwarzschild-Strasse 2, 85748 Garching, Germany\\
$^{13}$Centre for Astrophysics Research, Department of Physics, Astronomy and Mathematics, University of Hertfordshire, Hatfield AL10 9AB, UK\\
$^{14}$Steward Observatory, University of Arizona, 933 N. Cherry Avenue, Tucson, AZ 85721, USA\\
$^{15}$AURA for European Space Agency, Space Telescope Science Institute, 3700 San Martin Drive. Baltimore, MD, 21210\\
$^{16}$Department of Astronomy, University of Wisconsin-Madison, 475 N. Charter St., Madison, WI 53706 USA\\
$^{17}$Department of Astronomy and Astrophysics University of California, Santa Cruz, 1156 High Street, Santa Cruz CA 96054, USA\\
$^{18}$Max-Planck-Institut f\"ur extraterrestrische Physik (MPE), Gie{\ss}enbachstra{\ss}e 1, 85748 Garching, Germany\\
$^{19}$NSF National Optical-Infrared Astronomy Research Laboratory, 950 North Cherry Avenue, Tucson, AZ 85719, USA\\
$^{20}$NRC Herzberg, 5071 West Saanich Rd, Victoria, BC V9E 2E7, Canada\\
$^{21}$Cosmic Dawn Center (DAWN), Copenhagen, Denmark\\
$^{22}$Niels Bohr Institute, University of Copenhagen, Jagtvej 128, DK-2200, Copenhagen, Denmark
}

\date{Accepted XXX. Received YYY; in original form ZZZ}

\pubyear{2024}

\begin{document}
\label{firstpage}
\pagerange{\pageref{firstpage}--\pageref{lastpage}}
\maketitle

\begin{abstract}
The UV continuum slope of galaxies, $\beta$, is a powerful diagnostic of the metallicity and ages of stars, nebular gas properties, dust content, and potentially the escape of Lyman continuum photons from galaxies. Understanding the redshift evolution of $\beta$ and its dependence on key galaxy properties can therefore shed light on galaxy evolution over cosmic time. In this study, we present $\beta$ measurements for 295 spectroscopically confirmed galaxies at $5.5<z<14.3$ selected primarily from JADES, where $\beta$ has been measured from high quality JWST NIRSpec/PRISM spectra. We find a median $\beta = -2.3$ across our full sample, and find mild increase in blueness of $\beta$ with increasing redshift and fainter UV magnitudes. Interestingly, we find evidence for the average $\beta$ at $z>9.5$ to begin to redden, deviating from the trend observed at $z<9.5$. By producing stacked spectra in bins of redshift and $\beta$, we derive trends between $\beta$ and dust attenuation, metallicity, ionization parameter, and stellar age indicators directly from spectra, finding a lack of dust attenuation to be the dominant driver of bluer $\beta$ values. We further report six galaxies with $\beta \leq -3.0$, which show a range of spectroscopic properties and signs of significant LyC photon leakage. Finally, we show that the redder $\beta$ values at $z>9.5$ may require rapid build-up of dust reservoirs in the very early Universe or a significant contribution from the nebular continuum emission to the observed UV spectra, with the nebular continuum fraction depending on the gas temperatures and densities. Our modeling shows that in the absence of dust, nebular emission at $T>15,000$\,K can reproduce the range of $\beta$ that we see in our sample. Higher gas temperatures driven by hot, massive stars can boost the fraction of nebular continuum emission, potentially explaining the observed $\beta$ values as well as bright UV magnitudes seen across galaxies at $z>10$.
\end{abstract}




\section{Introduction}
The rest-frame ultraviolet (UV) continuum slope of galaxy spectra, $\beta$, which is parameterized as $f_\lambda \propto \lambda^\beta$, is an important quantity that can be used as a diagnostic of the nature of stellar populations and physical and chemical properties of the nebular gas in galaxies, the age and metallicity of stars, and the dust content \citep{cal94}. Detailed insights about the evolution of $\beta$ as a function of redshift can be used to infer the metallicity evolution \citep[e.g.][]{Calabro2021}, the change in the dust content in star-forming galaxies \citep[e.g.][]{Bouwens2009, Wilkins2013} and potentially the escape fraction of Lyman continuum (LyC) photons, that is crucial for evaluating the contribution of galaxies to the cosmic reionization budget \citep[e.g.][]{Chisholm2022}. 

Before the launch of \emph{JWST}, pioneering work has been done in measuring the evolution of $\beta$ with redshift for statistically significant samples of galaxies into the epoch of reionization using deep \emph{Hubble Space Telescope (HST)} imaging \citep[e.g.][]{McLure2011, Dunlop2012, Finkelstein2012, Wilkins2013, Bouwens2014, Bhatawdekar2021, Tacchella2022}. One of the main aims of such studies has been to identify galaxies with extremely blue UV slopes ($\beta \lesssim -3.0$) at $z>6$, which would likely trace extremely metal-poor and dust-free galaxies that are also possibly leaking a significant fraction of LyC photons to drive reionization \citep[e.g.][]{Bolamperti2023}. Indeed, some evidence of the existence of such blue UV slopes was reported \citep[e.g.][]{Bouwens2010, Finkelstein2010}. However, subsequent studies showed that these very blue $\beta$ measurements may suffer from observational and sample selection biases and that the median UV slope of $z\sim7$ galaxies may be nearer to $\beta \sim-2.5$ \citep[e.g.][]{Dunlop2012, Finkelstein2012, Bhatawdekar2021}. Further theoretical studies have also shown that at extremely low metallicities, the fraction of the total UV continuum at 1500\,\AA\ dominated by the nebular continuum components increases, thereby reddening the $\beta$ value \citep[e.g.][]{Raiter2010, Cameron2024_Neb, Katz2024}.

With \emph{JWST}, it is now finally possible to push measurements of $\beta$ for statistical samples of galaxies out to $z>10$. There have already been a number of studies exploring the dependence of $\beta$ both on redshift as well as the absolute UV magnitudes (\muv) using deep NIRCam imaging data out to $z>9$ \citep[e.g.][]{top22,top24a, Bouwens2023a, nanayakkara23, Furtak2023, cul23, cul23b, Austin2024}. Interestingly, most of these studies find evidence for the existence of galaxies with very blue UV slopes ($\beta < -2.7$), with a trend showing that UV-fainter galaxies consistently having bluer $\beta$ values.

Although studies relying on photometric data alone can lead to the identification of large samples, and can inherently be more `complete' in terms of the UV magnitude selection, they have their limitations. Perhaps the biggest source of uncertainty in photometrically-driven studies is the accuracy of photometric redshifts, and therefore the purity of the sample selection, particularly at fainter UV magnitudes \citep[e.g.][]{Bouwens2023b}. Contamination from strong rest-frame UV emission lines can also affect the measured $\beta$ from broad-band photometry alone, particularly if strong \lya\ emission is present in one of the crucial bands used to infer $\beta$. Finally, based on photometry alone, there is an unavoidable need to employ spectral energy distribution (SED) modelling to both accurately measure $\beta$ over the appropriate wavelength range as well as infer the physical and chemical properties of galaxies, which is crucial for breaking degeneracies between various parameters (such as dust reddening, age and metallicity of the stellar populations, etc.), which are responsible for setting the various features seen in the rest-frame UV spectra \citep[e.g.][]{cul23b, top24a}.

Thanks to large spectroscopic programs also being carried out by \emph{JWST}, particularly using the lower resolution PRISM/CLEAR configuration of the Near Infrared Spectrograph (NIRSpec) instrument \citep{fer22, jak22}, which gives simultaneous coverage across observed wavelengths of $\sim0.6-5.3$\,$\mu$m, it is now possible to spectroscopically measure the rest-frame UV continuum slopes of a large number of galaxies at high redshifts \citep[e.g.][]{rb24, heintz24, Hayes2024}. These studies have already shown a steady increase in the blueness of $\beta$ with redshifts across a range of UV magnitudes, which have been interpreted as the systematic reduction in the dust content of galaxies at earlier epochs, in line with findings from photometric studies. Additionally, $\beta$ measurements for some of the highest redshift galaxies, such as $\beta = -2.3$ for JADES-GS-z14-0 at $z=14.32$ \citep{Carniani2024}, $\beta = -2.5$ for GHZ2 at $z=12.34$ \citep{Castellano2024} and $\beta=-2.4$ for GN-z11 at $z=10.6$ \citep{bun23} now offer a remarkable opportunity to study the stellar and gas conditions within these record-breaking galaxies that regulate the emerging UV continuum slopes.

A key advantage of using spectroscopy to infer $\beta$ is the ability to mask UV metal lines, and the availability of a plethora of other key diagnostics available in galaxy spectra that can further shed light on the underlying physical and chemical conditions in the galaxies that may be responsible for setting the UV continuum slope \citep[e.g.][]{Calabro2021}. As mentioned earlier, key galaxy properties such as the age and metallicity of stars and gas, as well as the gas temperatures and densities can be inferred reliably from spectra using diagnostics relying on emission line fluxes and ratios, which play an important role in setting the UV slope of galaxies. To therefore understand the physical conditions within galaxies that may give rise to any relationship between $\beta$ and redshift or UV magnitude, performing an analysis on spectroscopically confirmed galaxies with robust detection of the UV continuum and emission lines is key.

In this study, we use a sample of 295 spectroscopically confirmed galaxies at $z>5.5$, spanning a wide range of UV magnitudes, to investigate the observed $\beta$ in galaxies at the highest redshifts. A key aim is to study how the redshift evolution of these properties could be used to explain the observed evolution in the UV continuum slopes of galaxies deep in the reionization era. We perform UV continuum slope measurements both for individual galaxy spectra as well as on stacked spectra constructed in carefully chosen bins in an attempt to inform both the median evolution of $\beta$, as well as what drives the scatter in its value over cosmic time, leveraging a number of diagnostics from galaxy spectra.

The layout of this paper is as follows. In Section \ref{sec:data} we describe the spectroscopic datasets and present the methodology used to perform measurements of quantities from galaxy spectra, including the $\beta$ using a novel Monte Carlo method. In Section \ref{sec: results} we present the main findings of our analysis, exploring the redshift evolution of $\beta$, its dependence on \muv, its correlation with galaxy physical and chemical properties using both individual measurements and stacking, as well as the discovery of a sample of galaxies showing $\beta \lesssim -3.0$. In Section \ref{sec:highz} we focus on $\beta$ measured for galaxies at $z>9.5$, highlighting the relatively redder $\beta$ measurements and its implications on the nature of some of the first galaxies that formed in the Universe. In Section \ref{sec:summary}, we summarize the key findings of this study.

Throughout this paper, we use the \citet{planck} cosmology. Magnitudes are in the AB system \citep{oke83} and all distances used are proper distances, unless otherwise stated. We adopt a solar metallicity of $Z_\odot = 0.02$.

\section{Data and measurements}
\label{sec:data}
\subsection{JADES spectroscopic data}
The datasets used in this work come from the \emph{JWST} Advanced Deep Extragalactic Survey (JADES; \citealt{bun20, eis23}), which has leveraged NIRSpec and NIRCam guaranteed time observations to obtain some of the deepest spectra and imaging of galaxies at high redshifts in the well-studied GOODS-South and North fields. The survey was designed such that `Medium' depth observations were obtained across both fields, with two additional `Deep' tier pointings observed in the GOODS-South field containing the footprint of the Hubble Ultra Deep Field (HUDF; \citealt{bun23b}). NIRSpec observations as part of JADES were obtained over a period of around 18 months, starting in October 2022 and ending in February 2024. Complementary NIRCam images in the fields were also obtained as part of JADES \citep{rie23}. For a full overview of JADES observations, survey design, coverage area, overlap with archival and other \emph{JWST} data, and survey depth and sensitivity, we refer the readers to \citet{eis23} and \citet{Deuginio2024}.

The NIRSpec observations in both fields across all tiers were carried out in Multi-Object Spectroscopy (MOS) mode \citep{fer22} in four filter/disperser combinations. These were the lowest resolution ($R\sim30-300$) but highest sensitivity PRISM/CLEAR configuration covering the full wavelength range from $\approx 0.6-5.3\,\mu\rm{m}$, along with three intermediate spectral resolution observations in G140M/F070LP, G235M/F170LP and G395M/F290LP giving $R\sim1000$ spectra across the full NIRSpec wavelength coverage. At the time of writing, the latest JADES data release is described in \citet{Deuginio2024}, which is Data Release 3.

For all practical purposes, the PRISM/CLEAR observations maximize the chances of detection of the continuum and emission lines, and the medium-resolution gratings' observations further enhance the accuracy of the measured redshift and provide more kinematic information and help separate key multiple emission lines that otherwise appear to be blended in PRISM spectra. Since the main aim of this study is to measure UV continuum slopes directly from the spectra of high redshift galaxies, we primarily use the $R\sim100$ PRISM/CLEAR spectra across all tiers of JADES where the UV continuum flux is detected with a decently high signal-to-noise ratio (S/N $>5$).

\subsection{Publicly available data}
In addition to JADES data, we also utilize the publicly available JWST Cycle 2 GO program 3215 (co-PIs: Eisenstein, Maiolino), which aimed to build upon JADES imaging and spectroscopy in the newly defined JADES Origin Field (JOF; \citealt{Eisenstein2023b}), which lies 8 arcminutes away from the HUDF in GOODS-S and was repeatedly observed as part of JADES observations in Cycles 1 and 2. We note here, however, that the NIRSpec MSA observations taken as part of PID 3215 cover the same region as in PID 1210, overlapping the HUDF.

The technical setup of the spectroscopic observations carried out as part of program 3215 is identical to the other JADES observations, with the same filter/disperser combinations being used. This enables a large degree of homogeneity between the JADES observations and those from program 3215.

\subsection{Spectroscopic redshift measurement}
Detailed information about all the steps involved in obtaining spectroscopic redshifts for our spectroscopic targets can be found in \citet{Deuginio2024}. However, in this section, we briefly describe the main steps involved in obtaining reliable spectroscopic redshifts. The redshift measurement procedure followed two steps. The first step involved obtaining a best-fitting spectral model to the observed spectrum using \texttt{BAGPIPES} \citep{car18} on the PRISM data. Armed with the \texttt{BAGPIPES} redshift, each spectrum was then visually inspected by at least two members of the JADES team, particularly focusing on the simultaneous detection of two or more strong emission lines at the same redshift.

Additional input was also taken from the higher resolution grating spectra to further refine both the line detection as well as redshift determination for all sources where emission lines were detected in the grating spectra. Based on the detection (or lack thereof) in both the PRISM and the grating data, a number of quality and reliability flags were then assigned to each source, which are described in \citet{Deuginio2024}. In particular, flags 6, 7 and 8 were used to identify `robust' redshift measurements thanks to the detection of multiple lines, which is what we largely use in this study. At the highest redshifts (e.g. $z>9.5$) when most bright rest-frame optical emission lines redshift out of the NIRSpec coverage, a redshift was determined mainly using the \lya-break feature in the continua. However, for galaxies at these high redshifts, additional insights on the validity of their spectroscopic redshift are obtained from multi-band NIRCam imaging and spectral energy distribution (SED) fitting, which effectively rules out these sources being low redshift interlopers \citep[e.g.][]{Hainline2024}.

\subsection{Final sample selection}
We note that the selection function of JADES, as is the case with the vast majority of NIRSpec programs utilizing the micro-shutter assembly (MSA) can be inhomogeneous across redshifts. This is mainly because of the way that shutters are allocated to higher-priority targets and the need to `protect' these high-priority targets from spectral overlap originating from other nearby sources. This exercise naturally results in the highest priority class sources (for example, at the highest redshifts) being more `complete' than sources with lower priority classes, which may be excluded from shutters to avoid contamination. 

For JADES, as described in \citet{bun23b}, sources at $z\gtrsim5.5$ ($i$-band dropout sources) were assigned systematically higher priorities than those at lower redshifts, resulting in a much higher fraction of such candidates ending up on MSA shutters. Sources at these redshifts are all mainly selected to be blue, star-forming galaxies (with the exception of a few `oddball'-type objects with redder spectra), which results in further sample homogeneity favoring star-forming galaxies at high redshifts compared to lower redshifts, where a wider variety of galaxy types were placed in shutters. 

The main driver, therefore, for the selection of Lyman-break galaxies at $z>5.5$ is the flux-limited nature of the survey, which manifests itself as a UV magnitude selection as a function of redshift. At progressively higher redshifts, the probability of photometrically-selected high-$z$ drop-out sources being placed in shutters increases, making JADES a highly targeted high-redshift galaxy survey \citep{bun23b}. Therefore, to leverage relatively `complete' samples for the analysis of rest-frame UV slopes of galaxies in this paper, we restrict our sample to galaxies with spectroscopic redshifts $z>5.5$, which will result in a more homogeneous and complete sample of Lyman-break selected galaxies that generally trace star-forming systems. We note, however, that with MSA surveys it may not be possible to obtain spectra for fully complete flux-limited samples. 

We also note that although a number of (candidate) AGN have been identified from the JADES datasets \citep[e.g.][]{Maiolino2023a, Scholtz2023, Lyu2024}, we do not explicitly exclude any such source as these numbers remain low compared to the number of star-forming galaxies, and particularly for low-luminosity AGN, the rest-frame UV light may still be dominated by young stars and/or nebular continuum emission.

\begin{table}
    \centering
    \caption{Number of $z>5.5$ sources with reliable (error $<5\%$) UV slope measurements belonging to different tiers/programs of JADES observations used in this study.}
    \begin{tabular}{l c c}
    \toprule
    Tier &  PID & N  \\
    \midrule 
    GS Deep/HST & 1210 & 32 \\
    GS Deep/JWST & 1287 & 22 \\
    GS-3215 & 3215 & 18 \\
    GS Medium/HST & 1180 & 40 \\
    GS Medium/JWST & 1180 & 22 \\
    GS Medium/JWST & 1286 & 80 \\
    GN Medium/HST & 1181 & 37 \\
    GN Medium/JWST & 1181 & 44 \\
    \midrule
    Total & & 295 \\
    \bottomrule
    \end{tabular}
    \label{tab:sample}
\end{table}

\subsection{Slit loss considerations}
An important observational effect that can potentially impact the measurement of rest-frame UV slope directly from spectra are the slit losses, which are nominally corrected for by the data reduction pipeline. Briefly, the point-spread function (PSF) of the instrument is wavelength dependent, and changes by a factor of $\sim3$ over the wavelength range $0.6-5.3$\,$\mu$m. Therefore, when producing the final calibrated science data products, the wavelength-dependent 2D light profile of the sources in the shutters needs to be taken into account and corrected for. The NIRSpec GTO data reduction pipeline takes the intra-shutter position of each target into account, which is used to compute the slit losses. The source is assumed to be a point source at the location of the input coordinates within the shutter, which works particularly well for high-redshift sources that are found to be generally small \citep[e.g.][]{Ormerod2024}.

To confirm the accuracy of $\beta$ measurements from spectra in light of possible slit loss related biases, we verify that the flux measurements across wavelengths in the spectra are compatible with photometric fluxes measured from NIRCam. In other words, we would like to confirm whether the continuum flux densities measured in NIRSpec spectra, which are prone to slit/shutter losses, are consistent with fluxes measured using circular apertures from the NIRCam images to identify and remove any biases in the $\beta$ measurements from spectra alone. 

To do this, we computed synthetic NIRCam photometry for all spectra in our sample, and compared it with the CIRC2 NIRCam photometry (a circular aperture of 0.3 arcsecond diameter, with an aperture correction applied appropriate for a point source) presented in \citet{rie23}. We found a remarkable agreement between the two flux measurements for the overwhelming majority of our sources, with outliers remaining few and far between. This finding confirms that (a) the slit loss corrections are largely being properly implemented in the data reduction and calibration pipeline and (b) the $\beta$ measurements directly from spectra are not being impacted by these slit losses that are unaccounted for. We note here that with the increasingly large PSF at longer wavelengths, any appreciable impact of slit losses will bias the measured UV slopes bluer, particularly at the highest redshifts.

\subsection{UV slope measurement}
To measure the UV slopes, $\beta$, from the PRISM/CLEAR $R\sim100$ spectra of all galaxies with robust spectroscopic redshifts, we implemented a novel Monte Carlo method that takes into account pixel-by-pixel uncertainties. The first step involved de-redshifting the galaxy spectra. We then proceeded to fit a power law to the rest-frame spectrum of the galaxies in the wavelength range $1340-2700$\,\AA, which is a slight modification of the standard \citet{cal94} windows to take into account the low spectral resolution of PRISM spectra that leads to broadening of the emission lines. This wavelength range avoids both the Lyman-alpha break and the Balmer break features in the continuum as well. To avoid fitting regions containing strong emission lines, we masked the regions $1440-1590$\,\AA, $1620-1680$\,\AA\ and $1860-1980$\,\AA. We note that the masked wavelength ranges are purposefully kept large to account for the broadening of the rest-UV lines due to the low resolution of the PRISM/CLEAR grating/disperser combination. We also note here that in the wavelength range used to measure $\beta$, the IGM damping wing feature should not play an important role in affecting $\beta$ unless the column densities are extremely high \citep[e.g.][]{heintz24}. 

Next, other outlying pixels (either due to other emission lines, residual cosmic rays or noise fluctuations) were masked out by using a sigma-clipping implemented at the $10\sigma$ level in the spectrum. Pixels with negative or zero flux were also masked, ensuring that the S/N of the neighboring pixels is always large enough such that the negative values only indicate bad data points.

Our Monte Carlo approach to measure the UV slopes was then implemented as follows. We performed 500 iterations and measured $\beta$ from each spectrum by fitting the power law function, where the flux in each spectral pixel was independently perturbed by randomly sampling from the distribution of the pixel-level noise. This approach produces a large range of possible $\beta$ values compatible with the pixel noise distribution. The median $\beta$ derived from the 500 iterations was set as the UV slope of the galaxy, with the standard deviation set as the $1\sigma$ error.
\begin{figure*}
    \centering
    \includegraphics[width=0.48\linewidth]{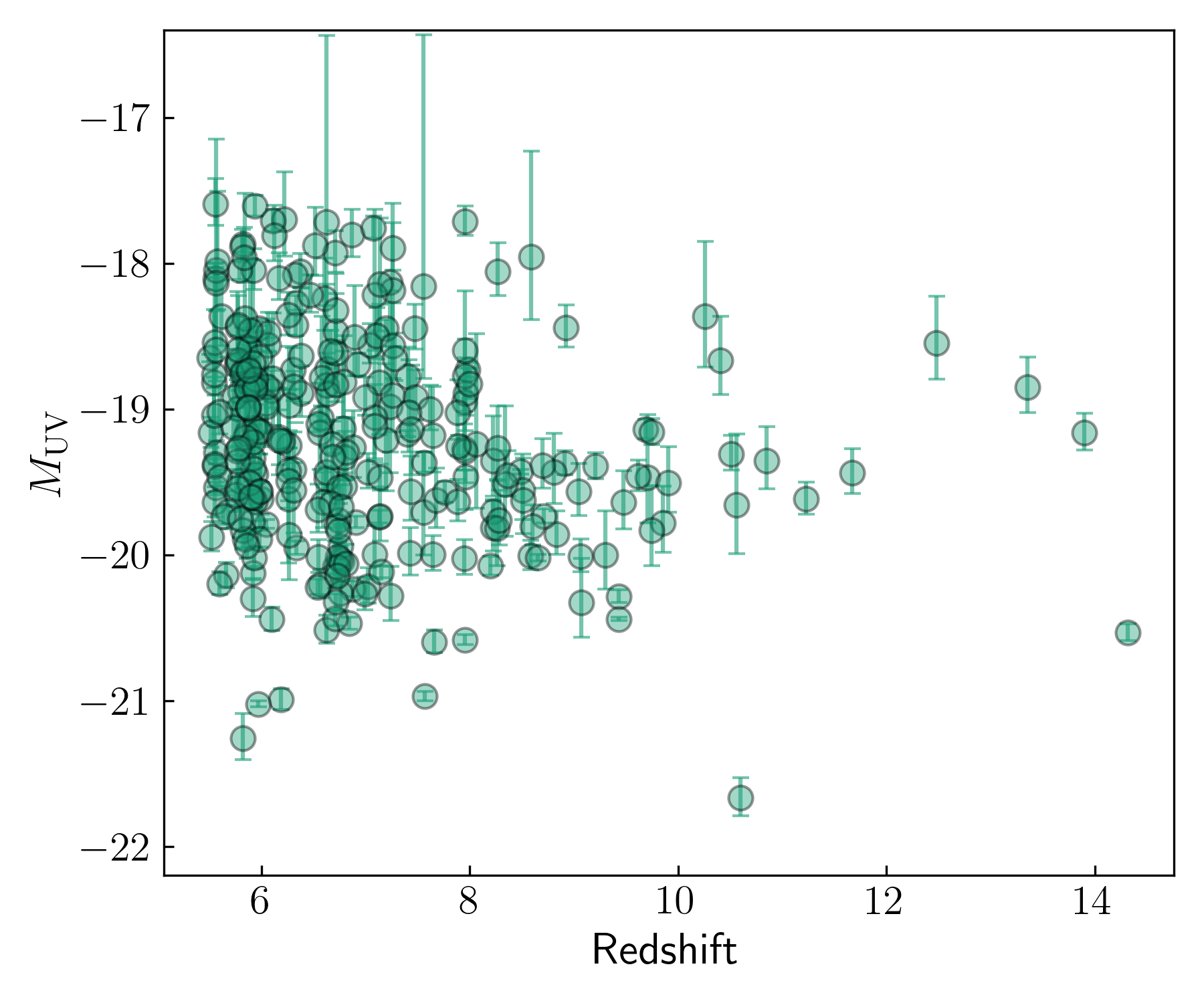}
    \includegraphics[width=0.48\linewidth]{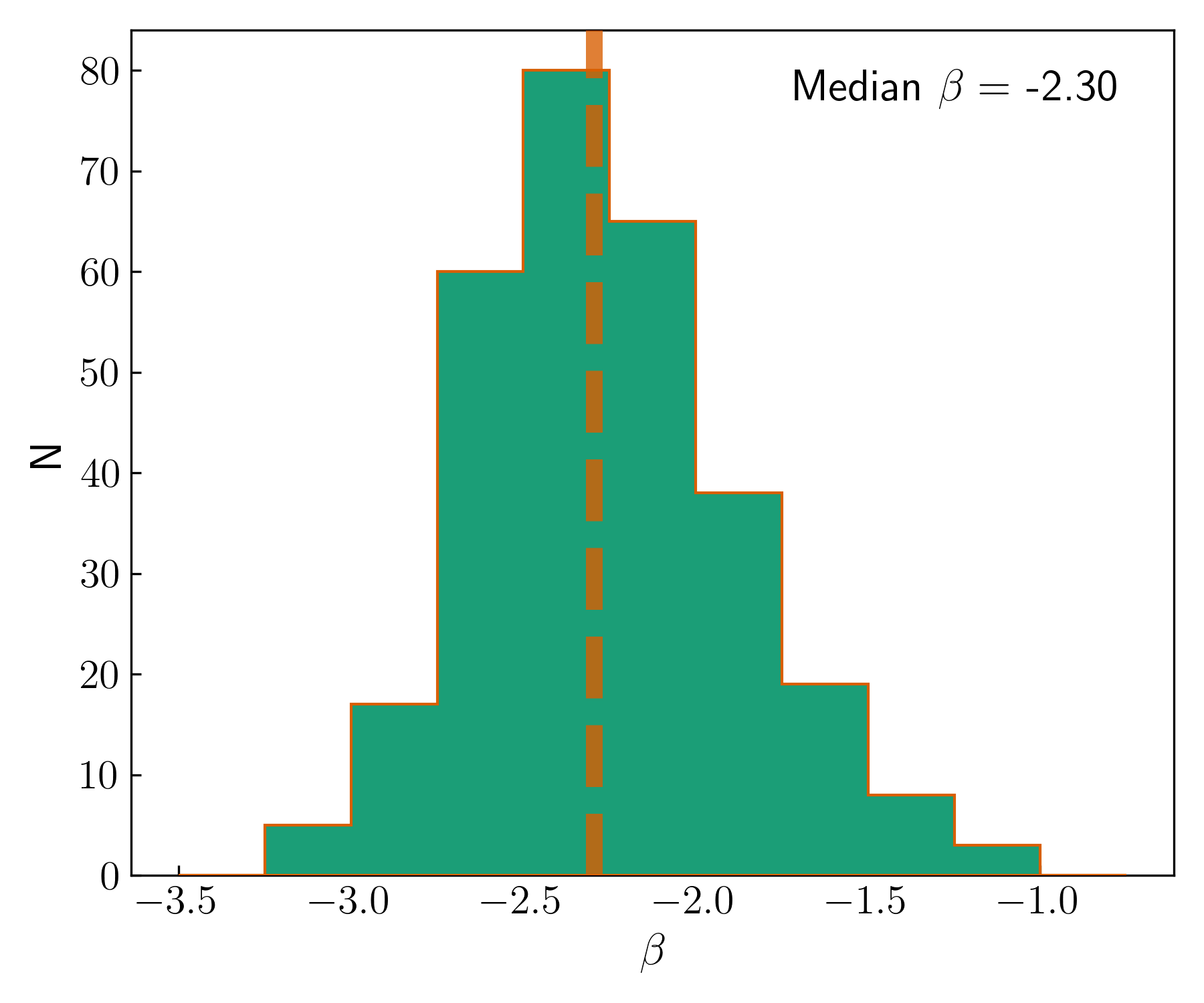}

    \caption{\emph{Left.} Overview of the spectroscopic sample considered in this study, which includes some of the highest redshift spectroscopically confirmed sources currently known. The continuum SNR\,$>2$ requirement to measure the UV slopes ensures that the UV magnitudes are robustly measured for all galaxies in our sample. \emph{Right.} Histogram of the UV slope $\beta$ of star-forming galaxies in this sample, with a median $\beta=-2.30$ measured across the entire sample spanning a large redshift range.}
    \label{fig:sample}
\end{figure*}

This $\beta$ measurement approach was implemented for all galaxies with robust spectroscopic redshifts in our sample. Given the variable depth of the various tiers of observations used in this study and the relatively broad redshift range resulting in variable UV absolute magnitude limits, it was deemed essential to weed out the rather uncertain $\beta$ measurements from our final sample before exploring their statistical distribution. Therefore, we opted to use an error threshold of $5\%$ to select our final sample corresponding to a $>2\sigma$ accuracy, which meant that galaxies with $\beta$ uncertainties larger than $5\%$ were rejected. Furthermore, less than $2\%$ of galaxies in our sample had a measured $\beta > -1$ with no strong nebular emission lines seen in the spectra, which may indicate that such galaxies are not actively forming stars and may represent other galaxy populations. In this work we wish to mainly study the properties of actively star-forming galaxies, and for that reason galaxies $\beta > -1$ were removed from our sample. The final sample across all tiers consisted of 295 galaxies, with robust redshifts and reliable $\beta$ measurements. A breakdown of the final sample across different tiers of observations is given in Table \ref{tab:sample}.

The distribution of the UV absolute magnitudes measured at rest-frame 1500\,\AA\ ($M_\mathrm{UV}$) and redshift of all galaxies for which $\beta$ could be reliably measured within the $5\%$ error threshold is shown in the left panel of Figure~\ref{fig:sample}. In the right panel of Figure~\ref{fig:sample} we show the histogram of $\beta$ measured across our sample, finding a median $\beta$ of $-2.30$.

\subsection{Binning and stacked spectra}
Although we can obtain a robust measurement of $\beta$ as well as physical and chemical properties of galaxies across the entire sample from individual spectroscopic measurements, we are particularly interested in investigating the dependence of $\beta$ on galaxy properties and the redshift evolution of UV slopes. To achieve this, we bin our sample in $\beta$ and redshifts, and produce stacked spectra weighted by S/N in each bin, where the S/N is measured at rest-frame $1500$\,\AA, to study how $\beta$ is related to the various physical and chemical properties traced by spectroscopic indicators. S/N weighting leverages the ultra-deep observations that were taken as part of JADES, which is able to achieve high S/N observations for relatively UV faint galaxies.

The binning in $\beta$ and redshift space is performed keeping the following points in mind. Since the redshift of the source plays an important role in dictating which emission lines will be visible in the observed NIRSpec spectrum, we wanted to ensure that redshift bins uniformly sample the observed strong emission lines, while ensuring that each redshift bin contained a comparable total number of sources. As a result, the lowest redshift Bin 1 covers the range $5.5 < z < 6$, Bin 2 $6<z<7$, where the cutoff at $z=7$ is chosen such that all sources in this bin have \ha\ visible in their spectra, Bin 3 $7< z < 8$, Bin 4 $8 < z < 9.5$, beyond which the \oiii+\hb\ complex moves out of NIRSpec coverage, and the highest redshift Bin 5 $z>9.5$.

\begin{figure}
    \centering
    \includegraphics[width=\linewidth]{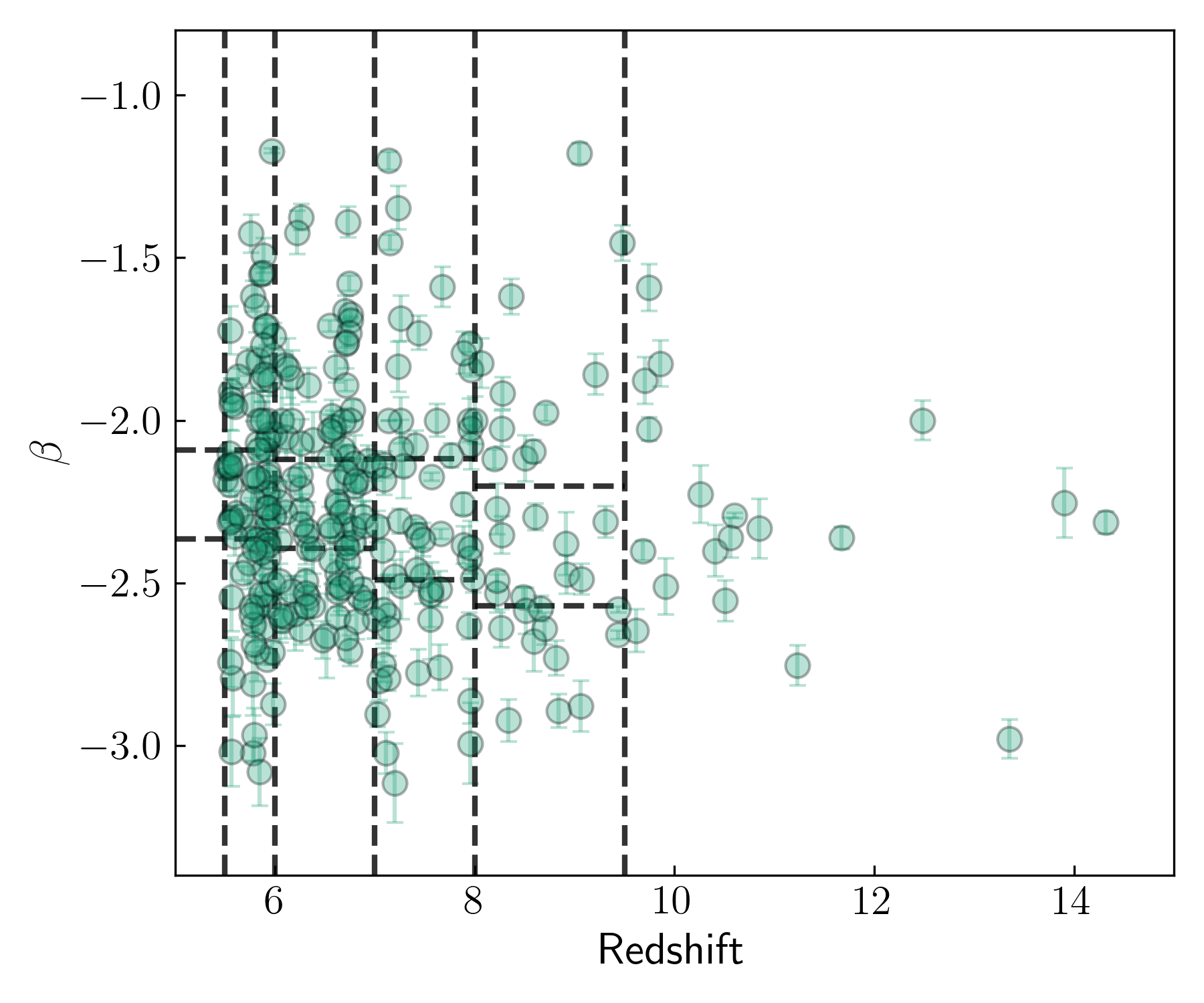}
    \caption{Binning of the sample in $\beta$ and redshift space, chosen to produce stacked spectra and study the sample averaged evolution of $\beta$ with redshift and other spectroscopic properties. The redshift bins were chosen to ensure coverage of key strong emission lines (as mentioned in the main text), whereas the $\beta$ bins were simply chosen to split the sample into three equally numbered tertiles. We do not split our $z>9.5$ sample into $\beta$ bins to preserve number statistics in the stack.}
    \label{fig:beta-redshift-bins}
\end{figure}

Within each redshift bin, we further split our sample into tertiles (three equally numbered bins) of $\beta$ to effectively obtain three baseline $\beta$ sub-samples across redshifts. The resulting sub-samples in $\beta$ are as follows: Bin 1A: $\beta > -2.10$, Bin 1B: $-2.10 > \beta > -2.36$ and Bin 1C: $\beta < -2.36$, Bin 2A: $\beta > -2.12$, Bin 2B: $-2.12 < \beta < -2.39$ and Bin 2C: $\beta < -2.39$, Bin 3A: $\beta > -2.12$, Bin 3B: $-2.12 < \beta < -2.49$ and Bin 3C: $\beta < -2.49$, and Bin 4A: $\beta > -2.20$, Bin 4B: $-2.20 < \beta < -2.57$ and Bin 4C: $\beta < -2.57$. We note here that due to the presence of only 19 galaxies in our highest redshift Bin 5, we do not split this in $\beta$ and treat the $z>9.5$ sample as a whole in this study. A visual representation of the bins in $\beta$-redshift space is shown in Figure \ref{fig:beta-redshift-bins}, with a breakdown of numbers in each subset of redshift and $\beta$ given in Table \ref{tab:subbins}.

\begin{table}
    \centering
    \caption{Sample splitting parameters in bins of redshift and $\beta$.}
    \begin{tabular}{l c c c}
    \toprule
    Bin & Redshift & Beta range & N \\
    \midrule
    1A & $5.5 - 6$ & $\beta > -2.10$ & 32 \\
    1B & $5.5 - 6$ & $-2.10 > \beta > -2.36$ & 30 \\
    1C & $5.5 - 6$ & $\beta < -2.36$ & 31 \\
    \midrule
    2A & $6 - 7$ & $\beta > -2.12$ & 32 \\
    2B & $6 - 7$ & $-2.12 > \beta > -2.39$ & 30 \\
    2C & $6 - 7$ & $\beta < -2.39$ & 31 \\ 
    \midrule
    3A & $7 - 8$ & $\beta > -2.12$ & 20 \\
    3B & $7 - 8$ & $-2.12 > \beta > -2.49$ & 19 \\
    3C & $7 - 8$ & $\beta < -2.49$ & 19 \\ 
    \midrule
    4A & $8 - 9.5$ & $\beta > -2.20$ & 11 \\
    4B & $8 - 9.5$ & $-2.20 > \beta > -2.57$ & 10 \\
    4C & $8 - 9.5$ & $\beta < -2.57$ & 11 \\ 
    \bottomrule
    \end{tabular}
    \label{tab:subbins}
\end{table}

We then produced stacked spectra in each bin in the following way. The spectra were first de-redshifted using accurate spectroscopic redshifts as mentioned earlier. The spectra in each bin were then resampled onto the same wavelength grid. Owing to our multi-tiered final sample containing spectra of variable depths and noise levels, we opt to use a SNR-weighted average stacking approach, which effectively up-weights spectra with higher SNR and down-weights spectra with lower SNR. We note that due to the multi-tiered construction of our sample, the SNR is not only dependent on the luminosity of the source, but also the depth of the tier of observations. Therefore, a SNR-weighted stacking scheme offers the best option to equally capture the contribution of galaxies occupying a range of UV fluxes/luminosities. The final weighted averaged spectra were then produced for each bin, with all the spectra shown in Figure \ref{fig:stacks1}.

For each stacked spectrum, we then performed a new measurement of $\beta$ following the same methodology as described earlier. We additionally measured  fluxes and ratios of key emission lines, which included \ha, \oiii\,$\lambda4959,5007$, \hb, \hg+\oiii\,$\lambda4363$ and \oii\,$\lambda\lambda3727,3729$, which appears to be blended. Using these lines, we calculated the dust attenuation from the Balmer decrements, the O32 (\oiii/\oii) ratio and the R23 (\oii+\oiii / \hb) ratio, from which information about the ionization parameters and metallicities could then be derived. The relevant measurements for all stacks are given in Table \ref{tab:stack_bins}.

\begin{table*}
    \centering
    \caption{Median properties measured from the stacked spectra of galaxies in bins of redshift and $\beta$. The Bin IDs are assigned as follows: Bins of increasing redshift are from Bin 1 to Bin 5, with each redshift bin split into 3 sub-bins of $\beta$ from reddest (A) to bluest(C). $^*$ denotes gas-phase metallicities measured using the strong line method \citep[e.g.][]{Curti2020} and $^\dagger$ denotes metallicities measured using the Direct $T_e$ method from the \oiii\,$\lambda4363$ auroral line detection.}
    \begin{tabular}{l c c c c c c c c c c c c}
    \toprule
    Bin & N & $z$ & $\beta$ & EW(\ha) & EW(\hb) & EW(\oiii$_{\lambda5007}$) & O32 & R23 &  E(B-V) & E(B-V) & 12+log(O/H)\\
    & & & & [\AA] & [\AA] & [\AA] & & & [\ha/\hb] &  [\hb/\hg] & \\
    \midrule
    1A & 32 & $5.822$ & $-1.81$ & $453.6 \pm 32.3$ & $63.3 \pm 4.7$ & $373.3 \pm 23.5$ & $5.4 \pm 0.1$ & $8.9 \pm 0.1$ & $0.23$ & $-$ & $8.19 \pm 0.10^*$ \\
    1B & 30 & $5.733$ & $-2.23$ & $761.4 \pm 54.0$ & $150.4 \pm 10.7$ & $846.8 \pm 59.9$ & $9.0 \pm 0.1$ & $8.1 \pm 0.1$ & $0.08$ & $-$ & $7.92 \pm 0.10^*$ \\
    1C & 31 & $5.814$ & $-2.63$ & $833.8 \pm 59.1$ & $133.6 \pm 9.6$ & $694.5 \pm 49.2$ & $9.9 \pm 0.6$ & $7.5 \pm 0.3$ & $0.12$ & $-$ & $7.84 \pm 0.15^*$ \\
    \midrule
    2A & 32 & $6.512$ & $-1.86$ & $586.7 \pm 42.1$ & $84.0 \pm 6.1$ & $466.5 \pm 33.0$ & $5.9 \pm 0.1$ & $8.2 \pm 0.2$ & $0.20$ & $0.34$ & $7.79 \pm 0.09^\dagger$ \\
    2B & 30 & $6.571$ & $-2.26$ & $1064.1 \pm 76.0$ & $106.3 \pm 7.7$ & $702.4 \pm 49.7$ & $11.0 \pm 0.3$ & $9.0 \pm 0.2$ & $0.20$ & $0.08$ & $7.77 \pm 0.09^\dagger$ \\
    2C & 31 & $6.496$ & $-2.55$ & $725.1 \pm 77.4$ & $159.2 \pm 11.4$ & $901.4 \pm 127.5$ & $9.4 \pm 0.1$ & $8.5 \pm 0.1$ & $0.13$ & $0.0$ & $8.05 \pm 0.09^\dagger$ \\
    \midrule
    3A & 20 & $7.563$ & $-1.83$ & $-$ & $168.3 \pm 13.0$ & $790.8 \pm 56.2$ & $5.9 \pm 0.1$ & $7.2 \pm 0.2$ & $-$ & $0.63$ & $7.70 \pm 0.09^\dagger$ \\
    3B & 19 & $7.478$ & $-2.34$ & $-$ & $167.1 \pm 12.5$ & $921.2 \pm 65.4$ & $13.5 \pm 0.5$ & $7.8 \pm 0.2$ & $-$ & $0.05$ & $7.78 \pm 0.09^\dagger$ \\
    3C & 19 & $7.393$ & $-2.73$ & $-$ & $365.6 \pm 28.9$ & $1657.7 \pm 118.1$ & $16.0 \pm 1.2$ & $6.4 \pm 0.4$ & $-$ & $0.0$ & $7.35 \pm 0.10^\dagger$ \\
    \midrule
    4A & 11 & $8.610$ & $-1.84$ & $-$ & $186.5 \pm 14.4$ & $1066.1 \pm 75.6$ & $8.1 \pm 0.3$ & $8.2 \pm 0.3$ & $-$ & $0.18$ & $7.70 \pm 0.09^\dagger$ \\
    4B & 10 & $8.624$ & $-2.41$ & $-$ & $146.1 \pm 11.8$ & $676.0 \pm 48.3$ & $12.2 \pm 0.6$ & $6.0 \pm 0.3$ & $-$ & $0.0$ & $7.30 \pm 0.09^\dagger$ \\
    4C & 11 & $8.785$ & $-2.71$ & $-$ & $83.4 \pm 9.3$ & $492.4 \pm 35.5$ & $8.6 \pm 0.6$ & $8.0 \pm 0.6$ & $-$ & $0.0$ & $7.79 \pm 0.10^\dagger$ \\
    \midrule
    5 & 19 & $10.969$ & $-2.30$ & $-$ & $-$ & $-$ & $-$ & $-$ & $-$ & $-$ & $-$ \\
    \bottomrule
    \end{tabular}
    \label{tab:stack_bins}
\end{table*}

\section{Results}
\label{sec: results}
Having measured $\beta$ for all galaxies in our sample, grouped them into bins created in the $\beta$ and redshift space, and created stacked spectra in each bin to study the dependence of $\beta$ on key galaxy spectroscopic properties, in this section we present some of the main findings of this study.

\subsection{The redshift evolution of the UV slope}
\begin{figure*}
    \centering
    \includegraphics[width=0.325\linewidth]{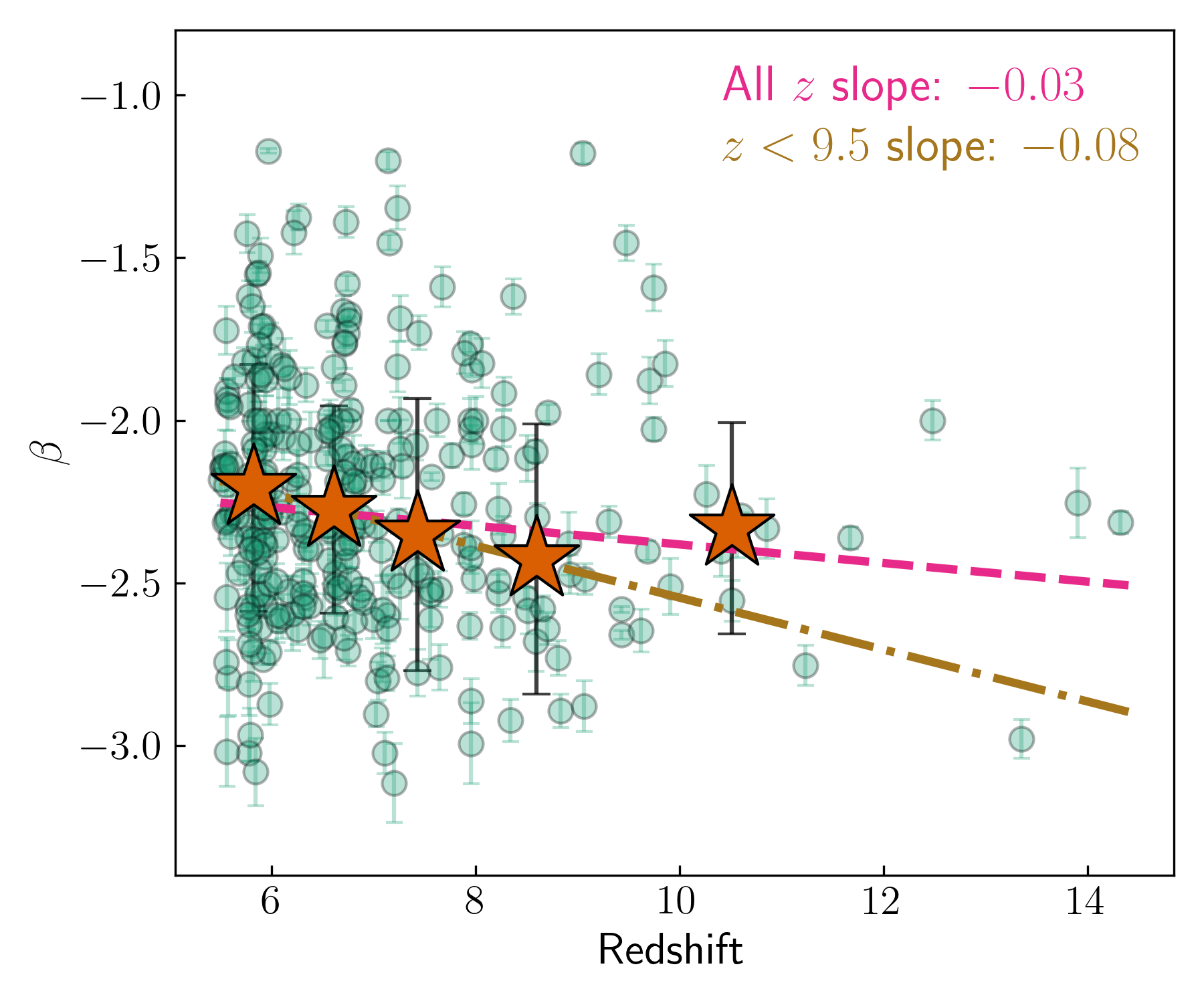}
    \includegraphics[width=0.325\linewidth]{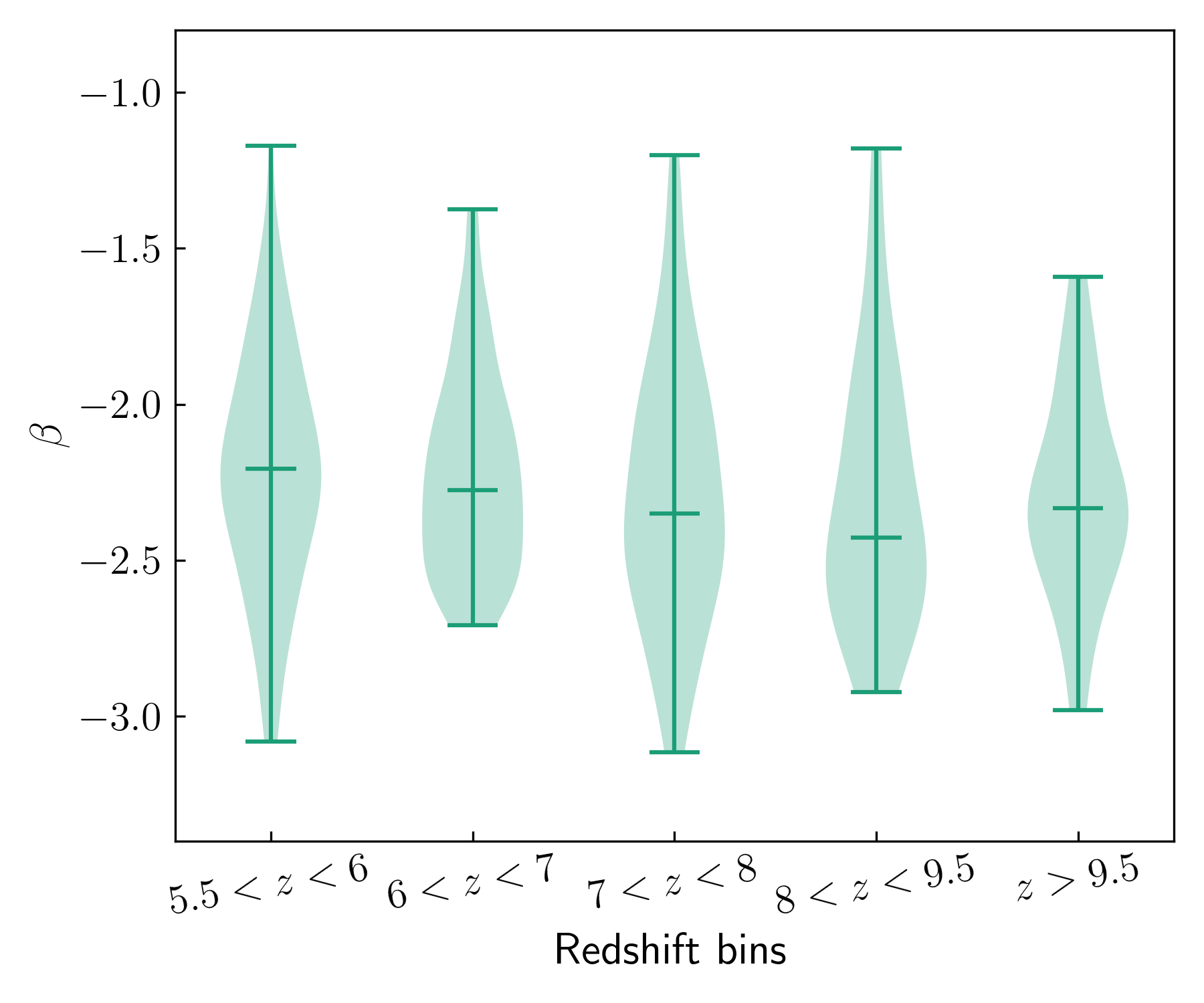}
    \includegraphics[width=0.325\linewidth]{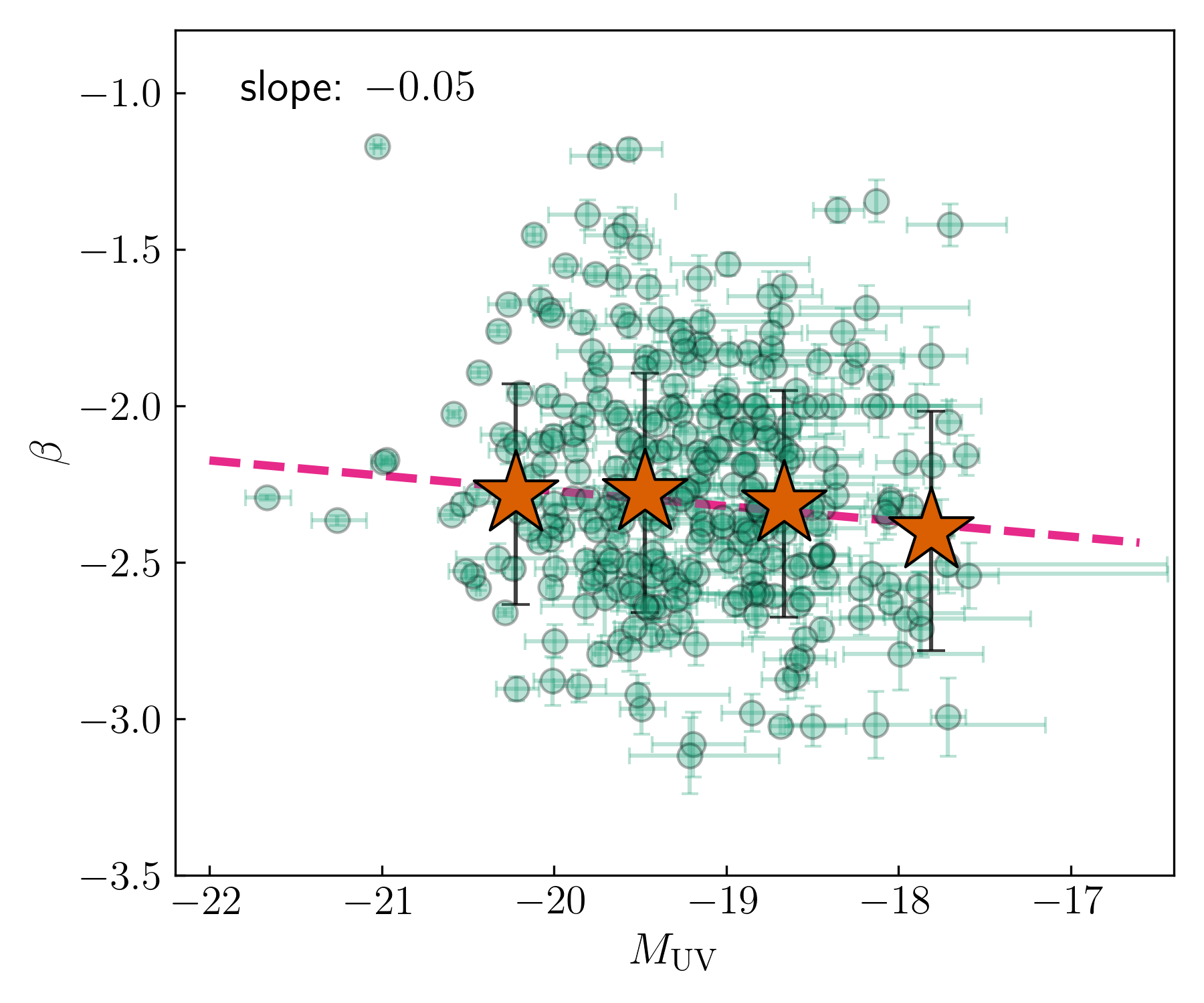}
    \caption{\emph{Left.} $\beta$ versus redshift, where the stars show the median and error bars show $1\sigma$ standard deviation in redshift bins. $\beta$ becomes bluer between redshifts $z\sim5.5-8$ but note that the $\beta$ values do not evolve much at $z\gtrsim8$. The median and standard deviation of $\beta$ measured in these redshift bins are given in the text. \emph{Middle.} Violin plots showing the median and standard deviation for the distribution of $\beta$ in the redshift bins considered in this study. We find that overall, the range of $\beta$ values is comparable across redshifts. At the highest redshifts, however, we do not see particularly red UV slopes ($\beta > -1.5$), but a slight reddening of the median $\beta$. \emph{Right.} Distribution of $\beta$ versus \muv\ for our sample, along with median $\beta$ values measured in \muv\ quartiles measured across the full sample. The median and standard deviation of $\beta$ measured in these \muv\ bins are given in the text. We find a very weak dependence of $\beta$ on \muv, with a shallow slope of the best-fitting relation of $-0.05$.}
    \label{fig:beta-redshift}
\end{figure*}

We begin by investigating the redshift evolution of $\beta$ across our spectroscopic sample. A plot of $\beta$ as a function of redshift is shown in the left panel of Figure \ref{fig:beta-redshift}, with both individual measurements and median values measured in the redshift bins 1 to 5. The median and standard deviation of $\beta$ that we measure in the redshift bins are as follows: Bin 1 ($z=5.82$): $\beta = -2.21 \pm 0.38$, Bin 2 ($z=6.61$): $\beta = -2.26 \pm 0.32$, Bin 3 ($z=7.43$): $\beta = -2.35 \pm 0.42$, Bin 4 ($z=8.60$): $\beta = -2.43 \pm 0.42$ and Bin 5 ($z=10.51$): $\beta = -2.33 \pm 0.32$. We find a very mild evolution of $\beta$ with redshift throughout the whole redshift range, with the best-fitting relation showing a relatively shallow slope of $-0.03$. The median values of $\beta$ decrease monotonically between redshifts 5.5 to $\sim7.5$, but show a slight upturn in the bin $z>9.5$.

To further explore this trend at $z>9.5$, we also derive the evolution of $\beta$ without Bin 5, finding a much bluer slope of the best-fitting relation of $-0.08$, as also shown in the left panel of Figure \ref{fig:beta-redshift}. This steeper trend would have predicted that most $z>9.5$ sources have much bluer $\beta$ values, and although some sources appear to align with this trend, the median value measured at $z>9.5$ is roughly in $1\sigma$ tension with the predicted trend at $z<9.5$. Although this tension is not strong, the apparent flattening of the relation between $\beta$ and redshift at $z>9.5$ is clear.

To further explore the distribution of $\beta$ values across redshifts, in the middle panel of Figure \ref{fig:beta-redshift} we show violin plots for the distribution of $\beta$ in the redshift bins considered in this study. It is clear that particularly in the $6<z<7$ and $8<z<9.5$ bins, we may be missing the tail of the distribution at the bluest $\beta$ values. This implies that in a more complete survey, the median $\beta$ measured at these redshifts may be even bluer. The $z>9.5$ bin on the other hand, despite suffering from low number statistics, appears to be a well-sampled normal distribution, giving further credibility to the idea of reddened $\beta$ at $z>9.5$. Lastly, we note that there are not many galaxies with $\beta < -3.0$ across the full redshift range, contrary to what has been reported by studies relying on photometric measurements of $\beta$ at these redshifts.

At $z>9.5$ in Bin 5, we find that the vast majority of galaxies have $\beta < -2$, consistent with stellar ages being relatively young, with recent star-formation and low dust content. We only see one galaxy with $\beta \sim -3$ at $z>9.5$, which is possibly tracing extremely young, stellar-light only dominated conditions with no dust attenuation. We will return to the properties of $z>9.5$ galaxies as inferred from their $\beta$ slopes and other spectroscopic indicators in the Discussion. 

Overall, we find a milder evolution of $\beta$ with redshifts when compared to the results from \citet{cul23b}. However, we note here that the \citet{cul23b} sample contained many more $z>10$ sources, although with only photometric redshifts. In particular, at $z>8$, we actually find a somewhat flipped trend compared to \citet{cul23b}. We note that at $z>10$, we do not have any $\beta$ measurements bluer than $-3$, whereas $\sim20$ sources in the \citet{cul23b} sample have $\beta < -3$. Galaxies with photometric $\beta < -3$ at $z>10$ were also reported in the photometric sample of \citet{cul23} and \citet{top24a}, which may play a role in enhancing the observed increase in blueness of $\beta$ with redshift. We also note that the typical uncertainty on the measurement of $\beta$ in our study is $\pm 0.1$, whereas for most photometric studies it is $\pm 0.4$.

\subsection{Dependence of the UV slope on UV magnitude}
Next, we investigate the dependence of $\beta$ on the UV magnitude of our galaxies. In the right panel of Figure \ref{fig:beta-redshift}, we show the distribution of $\beta$ with \muv, with median values measured in quartiles of \muv\ in our sample also shown. The median and standard deviation values $\beta$ that we measure in \muv\ bins -- from brighter to fainter -- for all redshifts are as follows: \muv\ bin 1 (\muv $=-20.22$): $\beta = -2.28 \pm 0.35$, \muv bin 2 (\muv $=-19.47$) = $\beta = -2.28 \pm 0.38$, \muv\ bin 3 (\muv $=-18.67$): $\beta = -2.31 \pm 0.36$ and \muv bin 4 (\muv $=-17.81$): $\beta = -2.40 \pm 0.38$. Overall, we find a slight dependence of $\beta$ on \muv\ with respect to the redshift, with $\beta$ becoming bluer at fainter \muv\ with a slope of $-0.05$. This \muv\ dependence is consistent with what has been reported from other photometric samples \citep[e.g.][]{cul23, cul23b, top24a}, in line with expectations that UV-fainter (or lower mass) galaxies must have fairly short, bursty, and recent star-formation histories and/or lower dust content \citet[e.g.][]{Narayanan2024}. Increased dust attenuation that may be expected at brighter \muv\ (or stellar masses) could also be playing a role in driving this \muv\ dependence of $\beta$.

\begin{figure*}
    \centering
    \includegraphics[width=\linewidth]{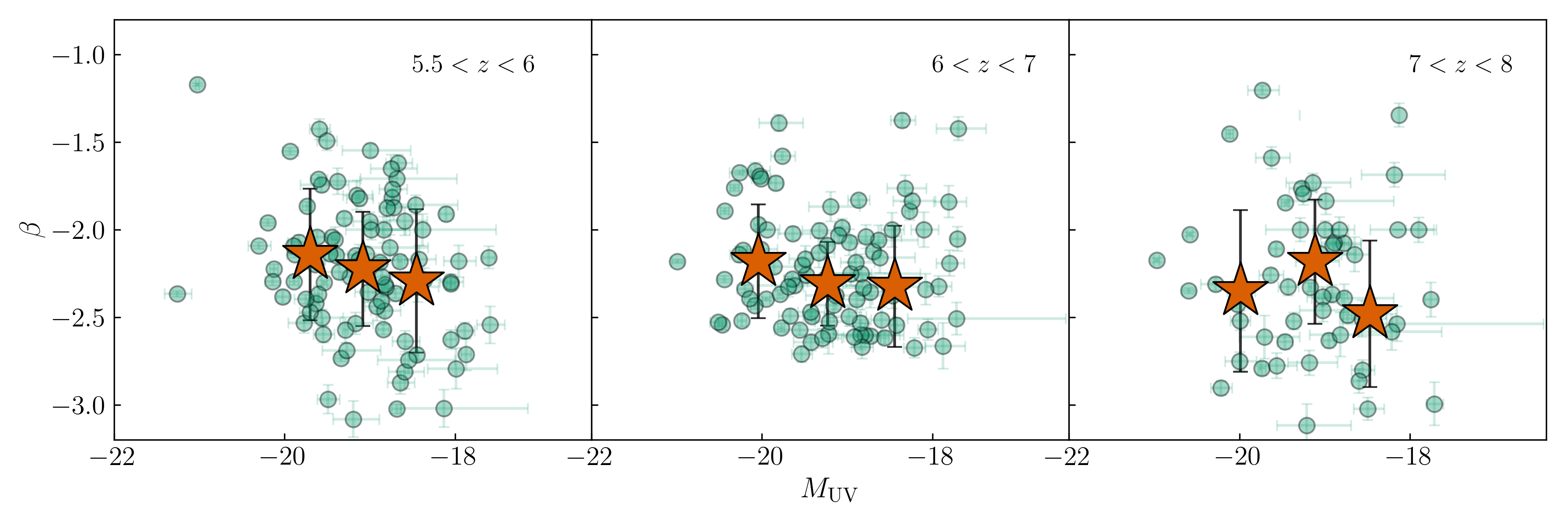}
    
    \includegraphics[width=0.66\linewidth]{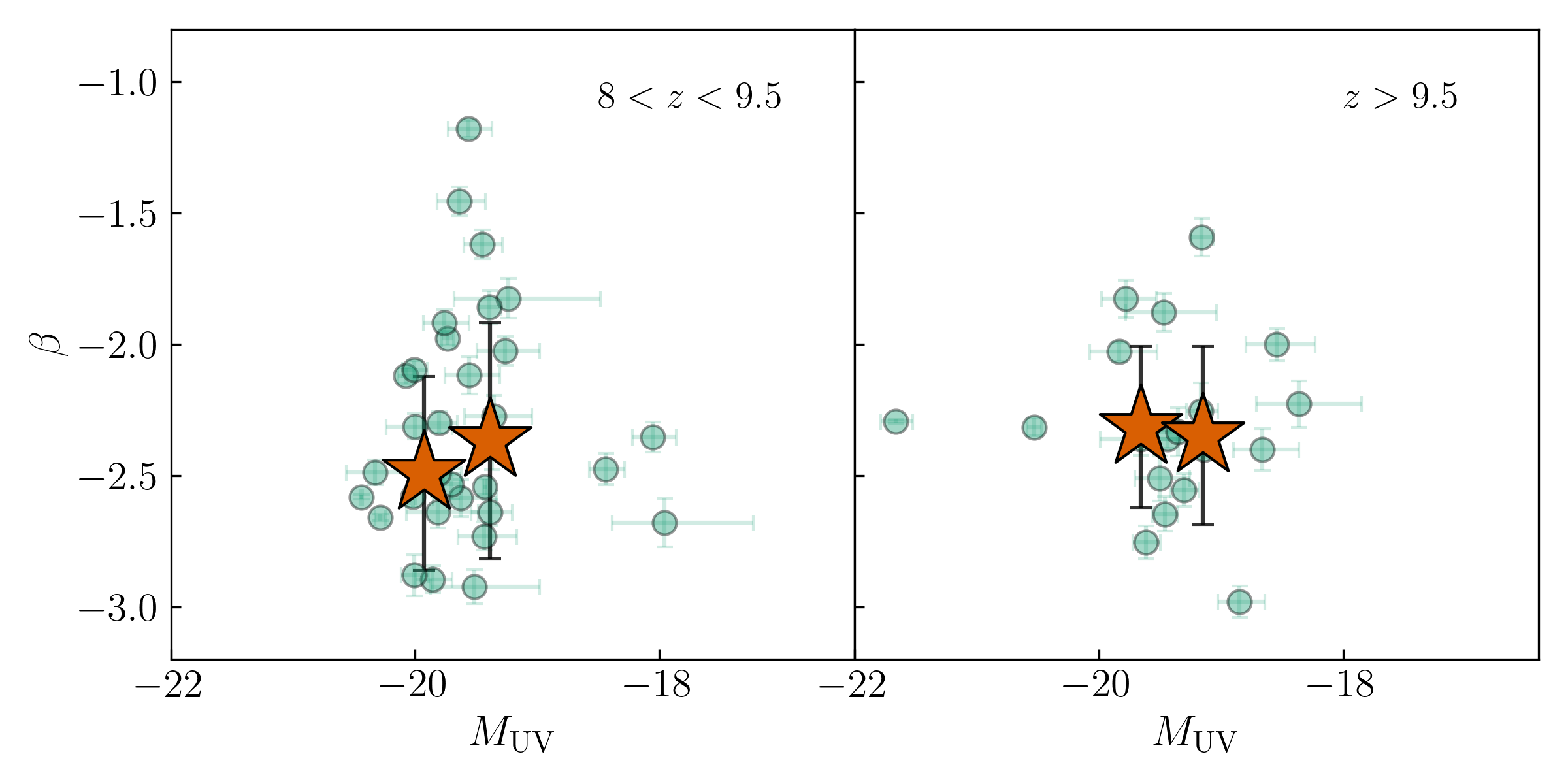}
    \caption{Dependence of $\beta$ on the galaxy UV magnitude in different redshift bins. The error bars represent the standard deviation in the bins. At redshifts $5.5<z<6$ and $6<z<7$ we do see a clear correlation between $\beta$ and UV magnitudes. However, there does not appear to be any dependence of $\beta$ on UV magnitude at $z>7$. This may be indicative of a lack of significant dust attenuation and its appreciable impact on the UV slopes of galaxies at $z>7$.}
    \label{fig:MUV-z-bins}
\end{figure*}

We further investigate the dependence of $\beta$ on \muv\ in our 5 independent redshift bins, as shown in Figure \ref{fig:MUV-z-bins}. To do this, we split the sample in redshift Bins 1, 2 and 3 into \muv\ tertiles (three equally populated bins), and Bins 4 and 5 into two sub-bins split by the median \muv. The subsets in \muv\ in each redshift bins are as follows: redshift Bin 1: \muv\ $< -19.40$, $-19.40 <$ \muv\ $< -18.77$ and \muv\ $>-18.77$, redshift Bin 2: \muv\ $< -19.63$, $-19.63 <$ \muv\ $< -18.84$ and \muv\ $>-18.84$, redshift Bin 3: \muv\ $< -19.44$, $-19.44 <$ \muv\ $< -18.85$ and \muv\ $>-18.85$, redshift Bin 4: \muv\ $<-19.63$ and \muv $>-19.63$, and redshift Bin 5: $<-19.43$ and \muv $>-19.43$.

In Bins 1 ($5.5<z<6$) and 2 ($6<z<7$), we see a very clear increase in blueness of $\beta$ at fainter \muv, indicating that either (a) low-luminosity galaxies have a high specific star-formation rate that has been shown to drive bluer $\beta$ values \citep{Narayanan2024}, and/or (b) high-luminosity galaxies have a higher dust content, thereby reddening the $\beta$ at the brighter end of the \muv\ distribution. \citet{Wilkins2013} indeed showed that increasing dust attenuation can explain the relationship between $\beta$ and \muv\ out to $z\sim7$, where the do see this relationship clearly in our sample.

However at $z>7$, we do not see clear trends between $\beta$ and \muv. We note here that the flux limited nature of our spectroscopic sample does lead to incompleteness at fainter \muv\ at the highest redshifts, which results in only a handful of galaxies fainter than \muv\,$>-18$ found at $z>7$. Regardless, the lack of dependence of $\beta$ on \muv\ could be attributed to the impact of dust attenuation on the UV slopes at $z>7$ may not be as significant as it may be at $z<7$, where dust has had time to be produced and grow.

The galaxies in the redshift bins at $z>7$ span a large range of $\beta$ values, ranging from $-1$ to $-3$. The \muv\ range at these redshifts is also relatively well sampled, although as noted earlier there is a missing population of galaxies with \muv\ fainter than $-18$. We also note that our $z>9.5$ bin contains two galaxies that are extremely UV-luminous with \muv\ brighter than $-20$, GS-z14-0 at $z=14.3$ and GN-z11 at $z=10.6$, which show relatively red $\beta$ values of $\beta \sim -2.3$.

In the next section, we investigate the dependence of $\beta$ on the physical and chemical properties of galaxies inferred directly from spectroscopy, to understand what physical conditions drive the observed relations of UV slope with redshift and \muv.

\subsection{UV slopes and galaxy physical properties}
\begin{figure*}
    \centering
    \includegraphics[width=0.95\linewidth]{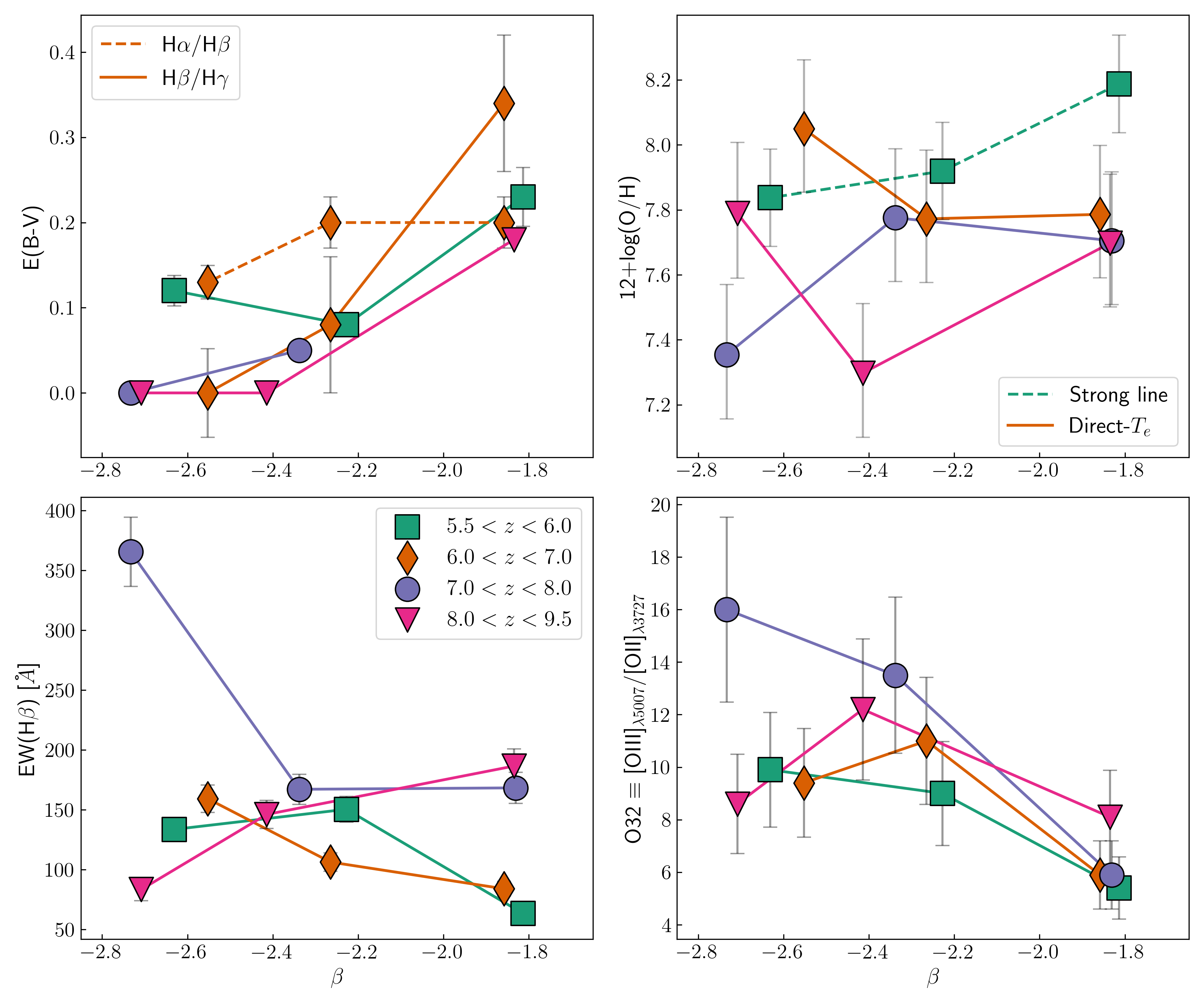}
    \caption{\emph{Clockwise from top left}: Dependence of galaxy properties such as dust attenuation or E(B-V) (top-left), gas-phase metallicity given by 12+log(O/H) (top-right), O32 ratio (bottom-right) and the \hb\ equivalent width (bottom-left) on the $\beta$ measured from stacked spectra in bins of redshift and $\beta$. The redshift bins are represented by different symbols and are used consistently across all panels. We find that dust attenuation measured using Balmer decrements of both \ha/\hb\ (dashed line) and \hb/\hg\ (solid line) decreases at bluer beta values, with little to no dust seen in the stacked spectra of our highest redshift bins. The metallicity, which we have measured using the strong-line method for the lowest redshift bins (dashed line( and direct-$T_e$ method (solid line) for the higher redshift bins, does not show any appreciable trend with $\beta$. The O32 ratio generally appears to increase at bluer $\beta$ values tracing younger stellar populations, but interestingly shows a downturn at the bluest $\beta$ values at the highest redshifts, which could be explained by a turbulent ISM in galaxies with the bluest UV slopes \citep[e.g.][]{Jin2022}. The \hb\ EW also does not show any clear trend with $\beta$.}
    \label{fig:beta-physical}
\end{figure*}

We now explore the properties inferred from the stacked spectra in our $\beta$ and redshift bins, and investigate the connection between $\beta$ and galaxy physical properties. We remind the readers that our sample was split into five redshift bins, Bins 1 to 5, and three $\beta$ bins, Bins A, B and C within each redshift bin (Table~\ref{tab:stack_bins}). The majority of the (rest-optical) emission line based properties that we present in this section are only possible for bins 1 to 4, as the highest redshift bin 5 does not cover rest-frame optical lines and the UV lines are not well-detected.

\subsubsection{Dependence on dust}
To measure the dust attenuation, we make use of the Balmer decrements using the \ha, \hb\ and H$\gamma$ lines whenever clearly visible in the stacked spectra. Owing to the variable spectral resolution of the NIRSpec PRISM, the \hg\ and \oiii$\lambda4363$ features appear to be heavily blended in the spectra at $z<7$. However, at these redshifts the \ha\ line is visible, and the dust attenuation at $z<7$ is calculated using the \ha\ and \hb\ lines. The \ha\ line moves out of NIRSpec coverage at $z>7$, which is also when the \hg+\oiii\,$\lambda4363$ feature becomes less blended due to an increase in resolution at the redder part of the observed spectrum. For these galaxies, we use the \hb\ and \hg\ ratios to infer dust attenuation. 

To calculate E(B-V) from the Balmer lines, we use the standard dust-free case-B recombination assumption with a temperature of $T_e = 10000$\,K and density $n_e = 300$\,cm$^{-3}$ \citep{ost06}. These assumptions give an intrinsic \ha/\hb\ ratio of $2.86$, and \hb/\hg\ ratio of $2.13$. We note here that given recent $T_e$ measurements from NIRSpec spectroscopy of galaxies at high redshifts \citep[e.g.][]{Laseter2024}, the assumption of $T_e = 10000$\,K may not be hot enough for such calculations, although this can be mitigated by the fact that the intrinsic value of the Balmer decrement does not change dramatically with temperature -- the \ha/\hb\ ratio at $T_e = 15000$\,K for example is $2.79$, representing only a $2.4\%$ change. Furthermore, it has been shown recently that Case-B recombination assumptions may not be valid across all high redshift galaxies \citep[e.g.][]{Scarlata2024, Yanagisawa2024, Mcclymont2024}, which will impact the amount of dust attenuation inferred from intrinsic ratios by up to $0.2$\,dex.

In the top-left panel of Figure \ref{fig:beta-physical}, we show the E(B-V) measured from Balmer lines for each $\beta$-$z$ bin as a function of $\beta$. Overall, we note a clear decrease in the dust content measured from stacked spectra with bluer $\beta$ values. These observed trends help confirm that an increased dust attenuation contributes towards making the $\beta$ less blue across all redshifts. 

\subsubsection{Dependence on gas-phase metallicity}
We now measure the metallicities from the stacked spectra in each bin using both the strong line method and the direct-$T_e$ method, which relies on the robust detection of the \oiii\,$\lambda4363$ line. As noted in the previous subsection, the \hg+\oiii\,$\lambda4363$ feature is heavily blended at $z\lesssim7$, preventing us from inferring the electron temperature from the \oiii\,$\lambda4363$/\oiii\,$\lambda5007$ ratio. At higher redshifts, however, the measurement of $T_e$ is easier. Therefore, for bins 1A, 1B and 1C, we infer the metallicity 12\,+\,log(O/H) using the strong line method, and for all other bins we employ the direct-$T_e$ method.

Strong line metallicities were inferred from the R23 ratio using the calibrations derived by \citet{Curti2020}. For the direct-$T_e$ method, we assume a density $n_e = 300$\,cm$^{-3}$ and derive the electron temperature $T_e$ using the grids provided in \textsc{pyneb} \citep{pyneb}. 

It is common to assume that the electron temperature of the region from which the \hb\ emission originates is the same as of \oiii, but certain scaling relations need to be used to estimate the temperature of the \oii\ emitting gas when the \oii\ auroral line is not clearly detected in the spectrum \citep[e.g.][]{Cameron2023b}. To estimate T(\oii), we use the scaling relations presented in \citet{Perez2017}, using the density dependent relation derived by \citet{Hagele2006} assuming $n_e = 1000$\,cm$^{-3}$.

In the top-right panel of Figure \ref{fig:beta-physical}, we show the interdependence between $\beta$ across redshifts and the gas-phase Oxygen abundance (measured as 12+log(O/H)) inferred from stacked spectra. As expected, we see a trend of decreasing metallicity at increasing redshifts, which has already been reported in the literature using \emph{JWST} spectroscopy \citep[e.g.][]{cur23, san23, nak23}. Interestingly, only in redshift bins 1 and 3 we see a clear decrease in metallicity with increasingly blue $\beta$. In the other bins, we do not see a clear trend, indicating that the gas-phase metallicity may not be a key driver of the observed $\beta$ in high redshift galaxies \citep[see also][for example]{curtis23}.

We note here that the $\beta$ arising from especially young, metal-poor stellar populations would require low stellar metallicities, which may not necessarily be reflected in the gas-phase metallicities. When the gas surrounding stars is enriched by the first generation of stars that formed out of chemically pristine gas, the gas gets polluted by heavier metals as soon as the first massive star winds and supernovae occur. Importantly, the gas-phase `metallicity' that we measure and report here is simply the oxygen abundance in the gas measured via emission lines. This is in contrast to the stellar metallicity, which usually reflects the level of iron enrichment among other heavier elements \citep[see][for example]{Cullen2021}. Therefore, the gas-phase oxygen abundance may not be a good tracer of the metallicity of the underlying stars that are producing the ionizing radiation and largely setting the $\beta$ values. Therefore, gas-phase metallicities would not be expected to correlate very strongly with $\beta$, which is what we find from our data. We do note that the gas-phase oxygen abundance appears to decrease with increasing redshifts across our sample \citep[see also][]{rb24}. However, in case of a significant contribution of nebular continuum emission, the observed UV slope may not necessarily correlate with the stellar metallicities either.

\subsubsection{Dependence on EW(\hb)}
The rest-frame equivalent width of the \hb\ line (along with other Balmer recombination lines) has been shown to be a good tracer of stellar ages, as the decrease in ionizing photon flux with increasing stellar age leads to the depletion of the nebular component of the Balmer emission lines \citep[e.g.][]{Leitherer2005, Levesque2013}. How the EW(\hb) actually decreases with increasing age, however, depends on a number of factors such as the metallicity and the dust content of \hii\ regions, stellar rotation, and whether massive stars are formed in binaries. Another important parameter that can reduce the observed \hb\ flux is the escape of Hydrogen ionizing Lyman continuum (LyC) photons.

In the bottom-left panel of Figure \ref{fig:beta-physical} we show the distribution of EW(\hb) measured from our stacked spectra with $\beta$. Once again, we find no clear trends in all of our redshift bins. At redshifts $5.5<z<8$, EW(\hb) appears to increase with a increasingly bluer $\beta$, indicating that galaxies with bluer UV slopes are likely powered by relatively younger stellar populations that are capable of producing a larger number of Hydrogen ionizing photons. 

The observed trends in the highest redshift bin with $8<z<9.5$ are interesting and somewhat unexpected. The EW(\hb) actually keeps decreasing with increasingly blue $\beta$, contrary to expectations from a younger stellar population being dominant at the highest redshifts in galaxies that also have the bluest UV slopes. This decrease in the ionization of nebular gas even at blue UV slopes may be attributed to increasingly non-zero escape fraction of LyC photons, which removes ionizing photons that would otherwise have ionized the Hydrogen gas in the galaxy \hii\ regions, depositing them instead in the galaxy CGM/IGM, driving the process of reionization \citep[e.g.][]{top22}.

\subsubsection{Dependence on O32 ratio}

Low metallicity conditions where young stars are forming generally result in higher ionization parameters that often lead to higher O32 ratios \citep[e.g.][]{Jaskot2013, paa18}. From measurements across our stacked spectra, we generally find that the O32 ratio increases at bluer $\beta$ values, as can be seen in the bottom-right panel of Figure \ref{fig:beta-physical}. However, we note that for the redshift $8-9.5$ bin, there is a decrease in the O32 ratio for the bluest beta slope bin, and the highest O32 ratio is seen in the intermediate beta bin.

One possible pathway to lower the O32 in the presence of a young, highly ionizing and dust-free stellar population is by altering the geometry of the \hii\ regions. \citet{Jin2022} showed that increased turbulence can decrease the O32 ratio along certain lines of sight when making spectroscopic observations by apparently boosting the relative contribution of the lower ionization emission lines such as \oii, originating from the outer parts of the \hii\ region \citep[see also][]{Katz2022Prism}. A turbulent ISM with a fractal-like geometry could increase the surface area covered by the `edges' of the ionization front, as opposed to what would be expected assuming a simple spherical geometry, which can increase the fractional volume of the \hii\ region that has a lower ionization state when observed at certain viewing angles. These differences are expected to be particularly pronounced in spatially resolved spectroscopy. Density and temperature fluctuations in the ISM have indeed been shown to change the inferred properties of the ISM from emission line diagnostics \citep[e.g.][]{Cameron2023} due to line of sight effects, and such effects are typically not captured by simple spherically symmetric models of emission from \hii\ regions.

It may also be possible that the low O32 ratio in the bluest $\beta$ bin at $8.0<z<9.5$ may be tracing non-turbulent, ionization-bounded \hii\ regions, as opposed to a density-bounded nebula that is typically seen in high redshift galaxies with strong underlying sources of ionization. In an ionization bounded scenario, the \oiii\ emission dominates closer to the ionizing source where the gas is in a higher ionization state, and the \oii\ emission dominates further away from the source where gas is in a lower ionization state. However, in ionization bounded nebulae with no ionizing photon escape, the Balmer emission lines arising from recombination (\ha, \hb) appear to be strong. As noted in the previous subsection, we also found that the EW(\hb) in the bluest $\beta$ bin at $8.0<z<9.5$ was low, which would perhaps require non-zero LyC \fesc. Therefore, in a non-turbulent ionization-bounded \hii\ region scenario, a morphology that has `holes' from which ionizing LyC photons may be able to escape would be needed to consistently explain both the low O32 as well as the low EW(\hb).

In many ways, the turbulent ISM model from \citet{Jin2022} effectively leads to a scenario that is more akin to being ionization-bounded by increasing the surface area where lower ionization gas exists, thereby driving down the observed O32 ratios. The turbulent geometry of the gas may naturally give rise to ionizing photon escape channels, thereby also explaining the lower observed EW(\hb). Furthermore, turbulence in the ISM may be expected in galaxies where a very recent or extreme starburst event has taken place, traced by the observed extremely blue UV slopes. It is possible that these galaxies are only just beginning to assemble the bulk of their stellar masses, and therefore, their ISM may be undergoing a highly turbulent phase.

\subsection{Extremely blue UV slopes: $\beta \leq -3.0$}
\begin{figure*}
    \centering
    \includegraphics[width=0.49\linewidth]{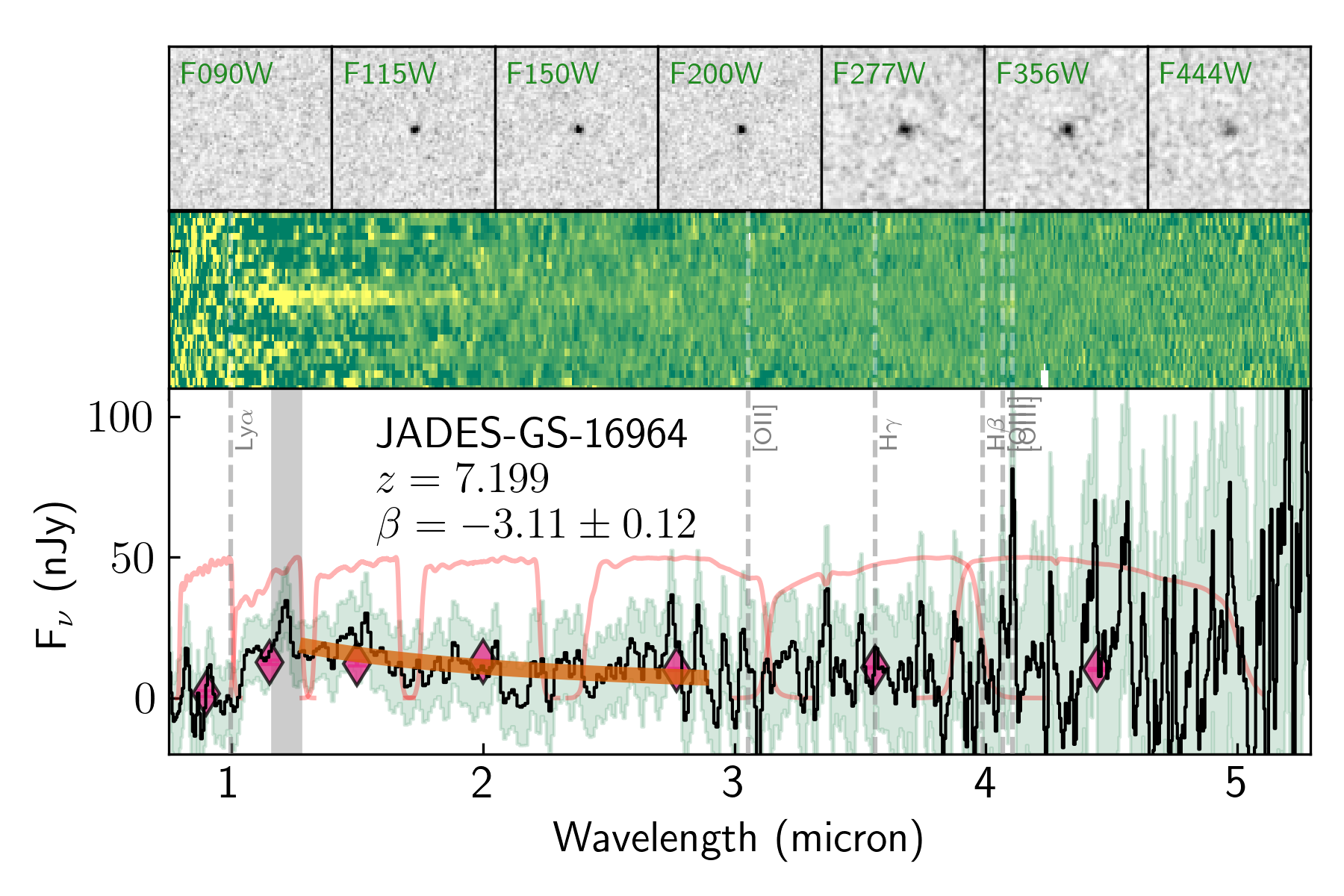}
    \includegraphics[width=0.49\linewidth]{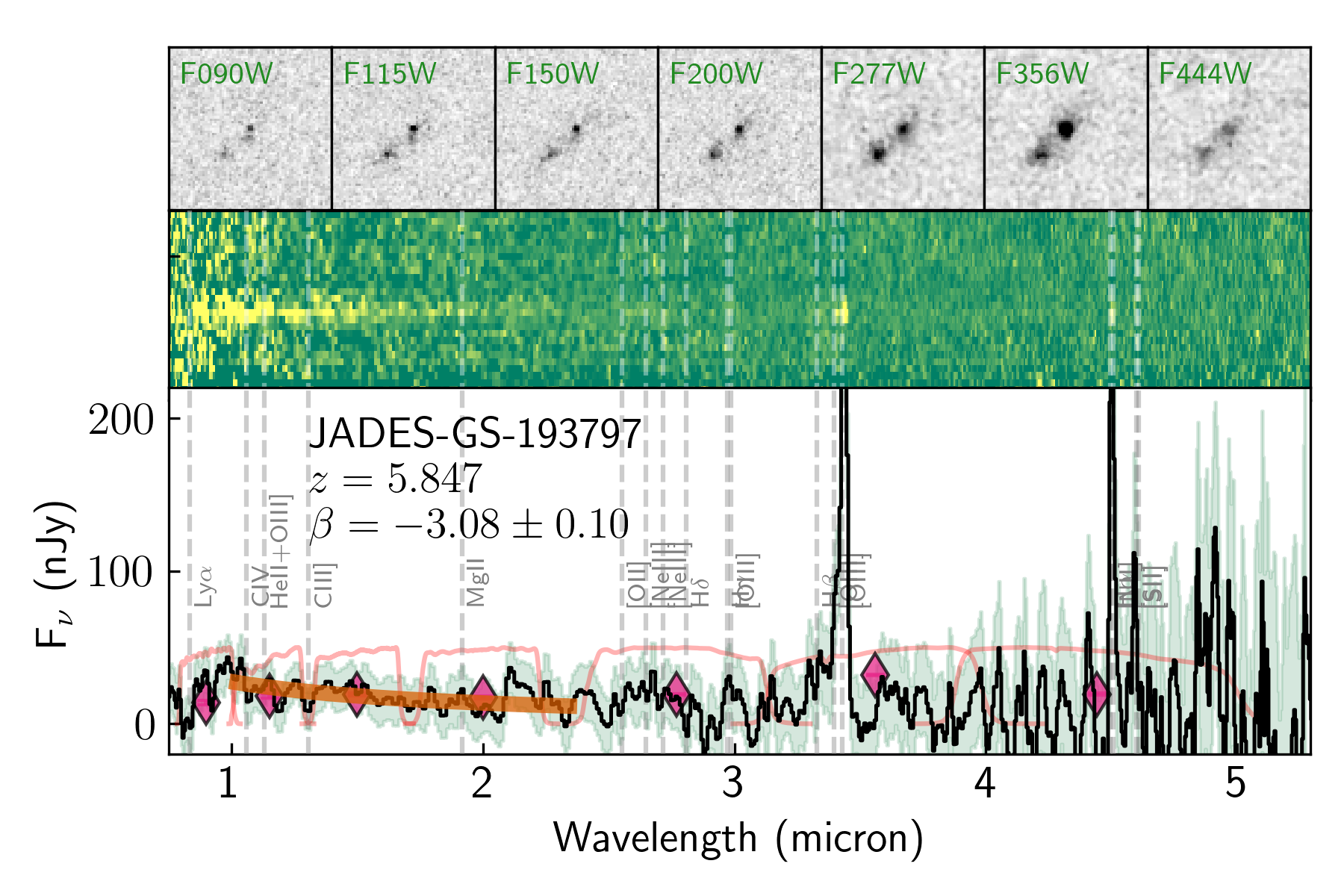}
    
    \includegraphics[width=0.49\linewidth]{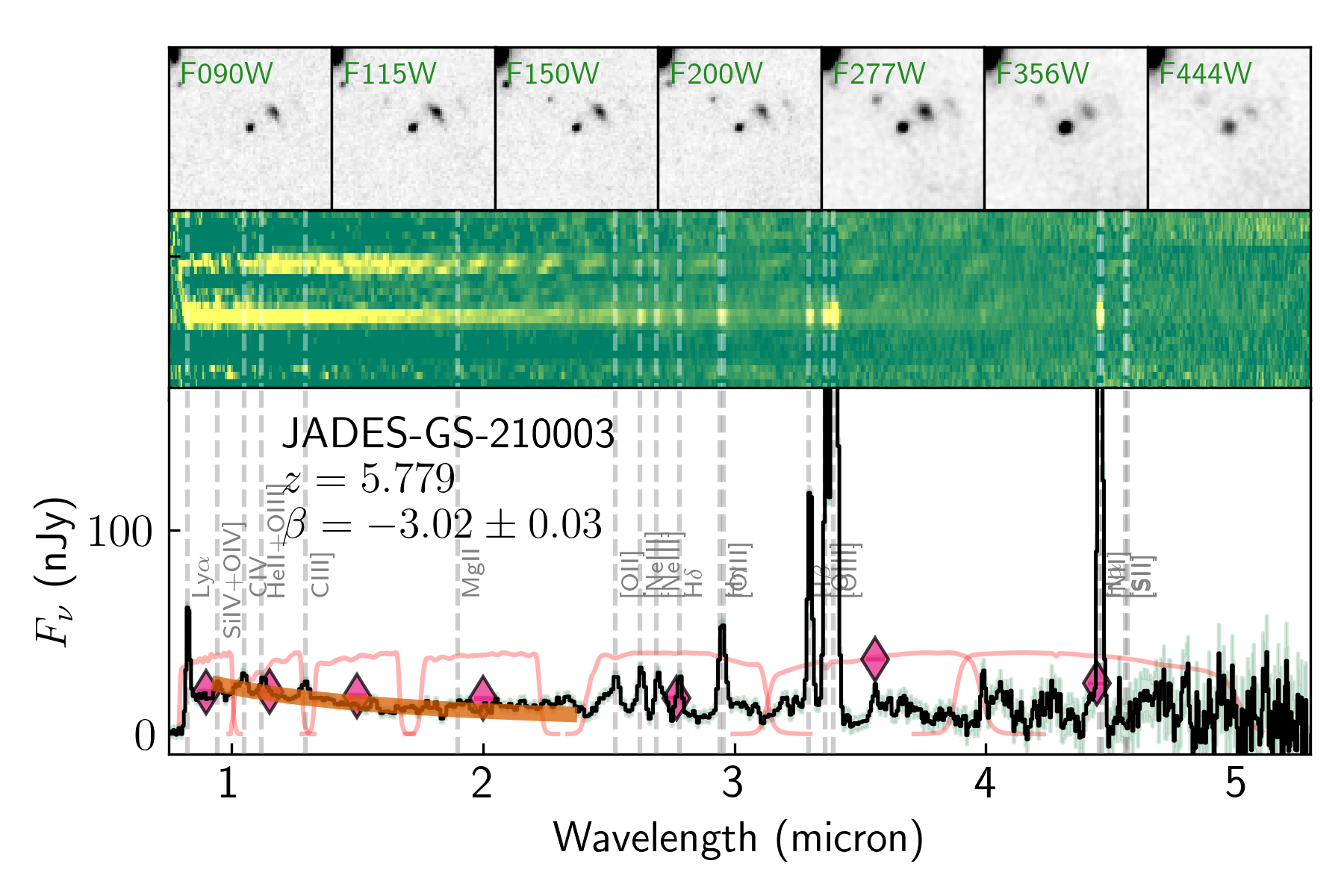}
    \includegraphics[width=0.49\linewidth]{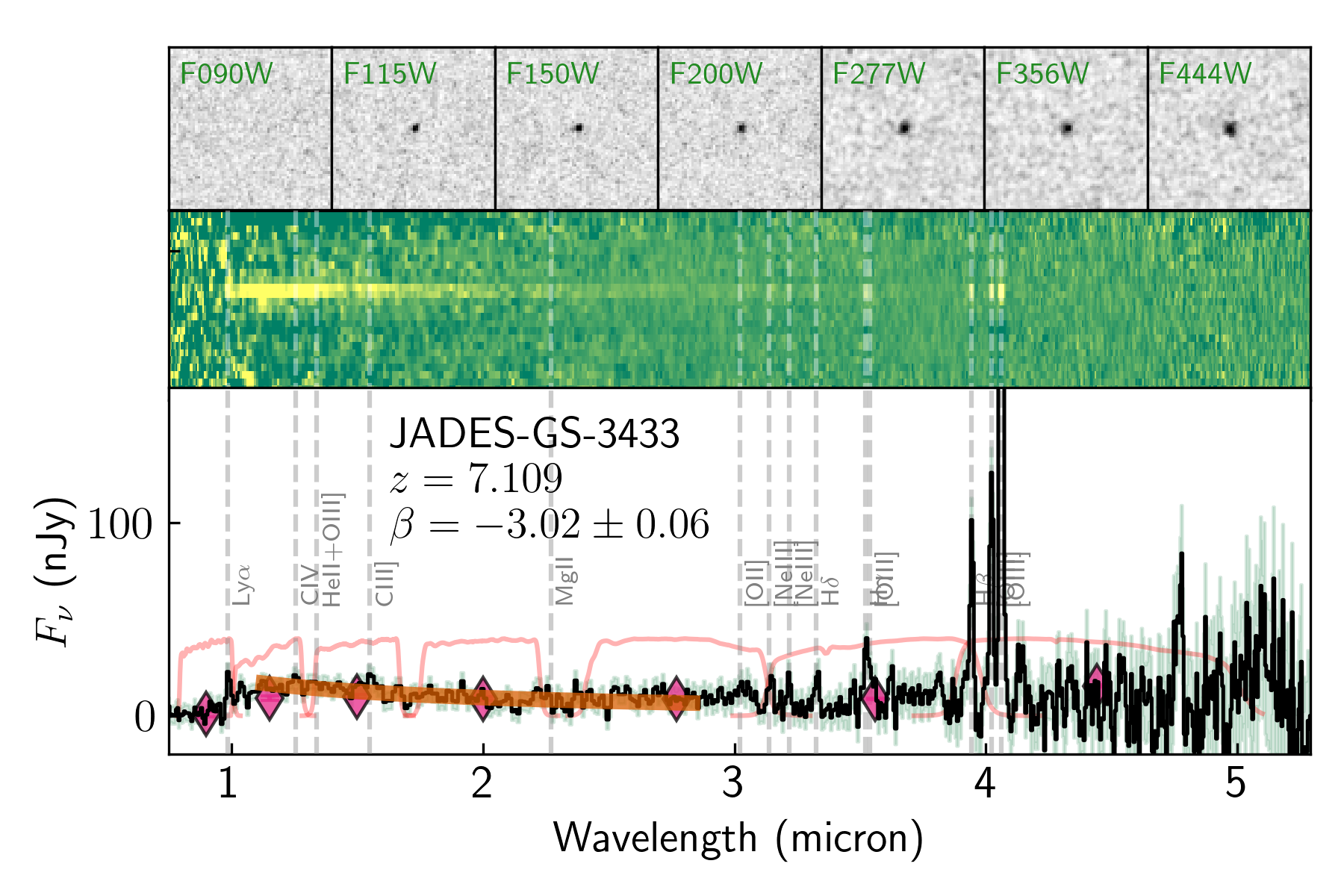}
    
    \includegraphics[width=0.49\linewidth]{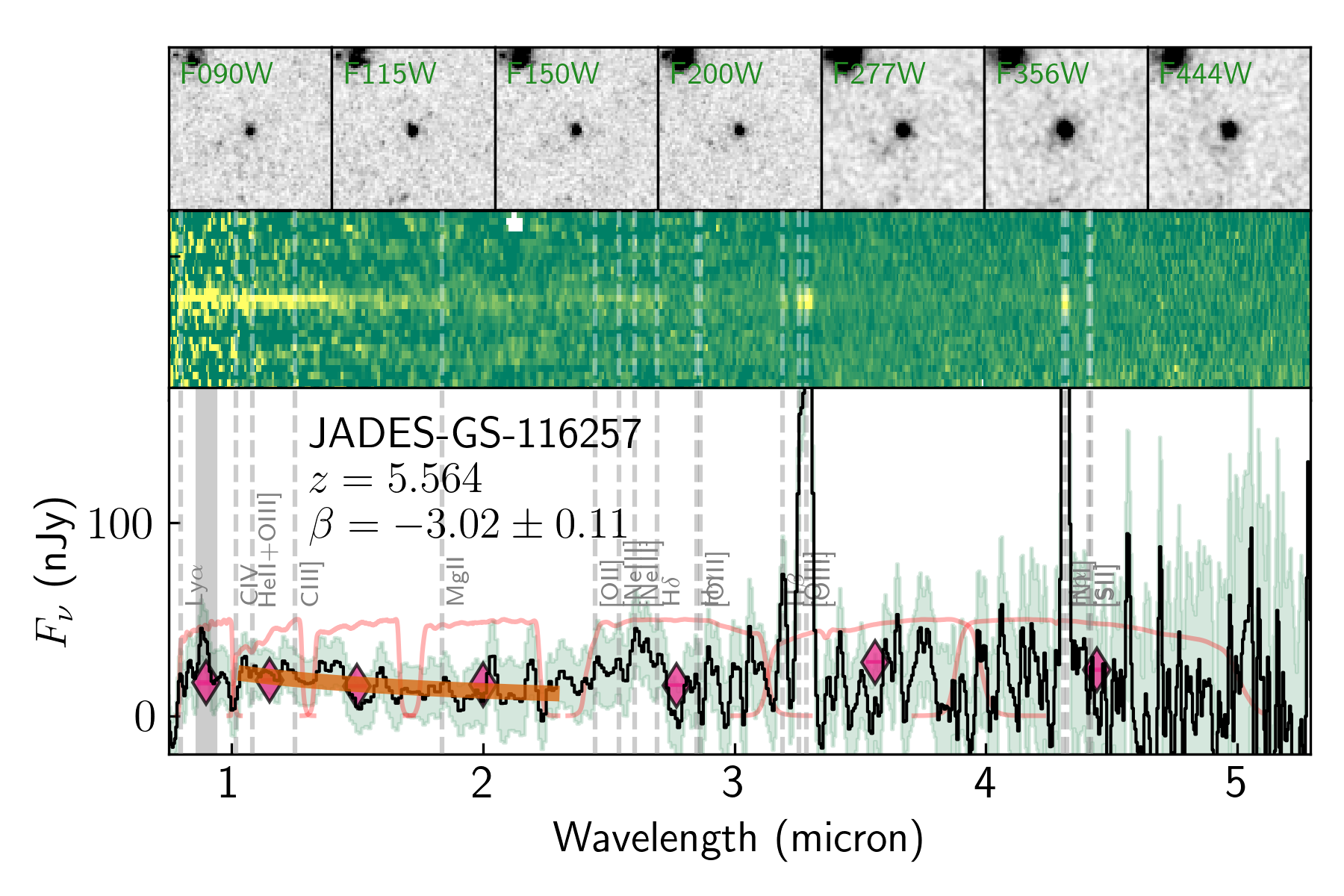}
    \includegraphics[width=0.49\linewidth]{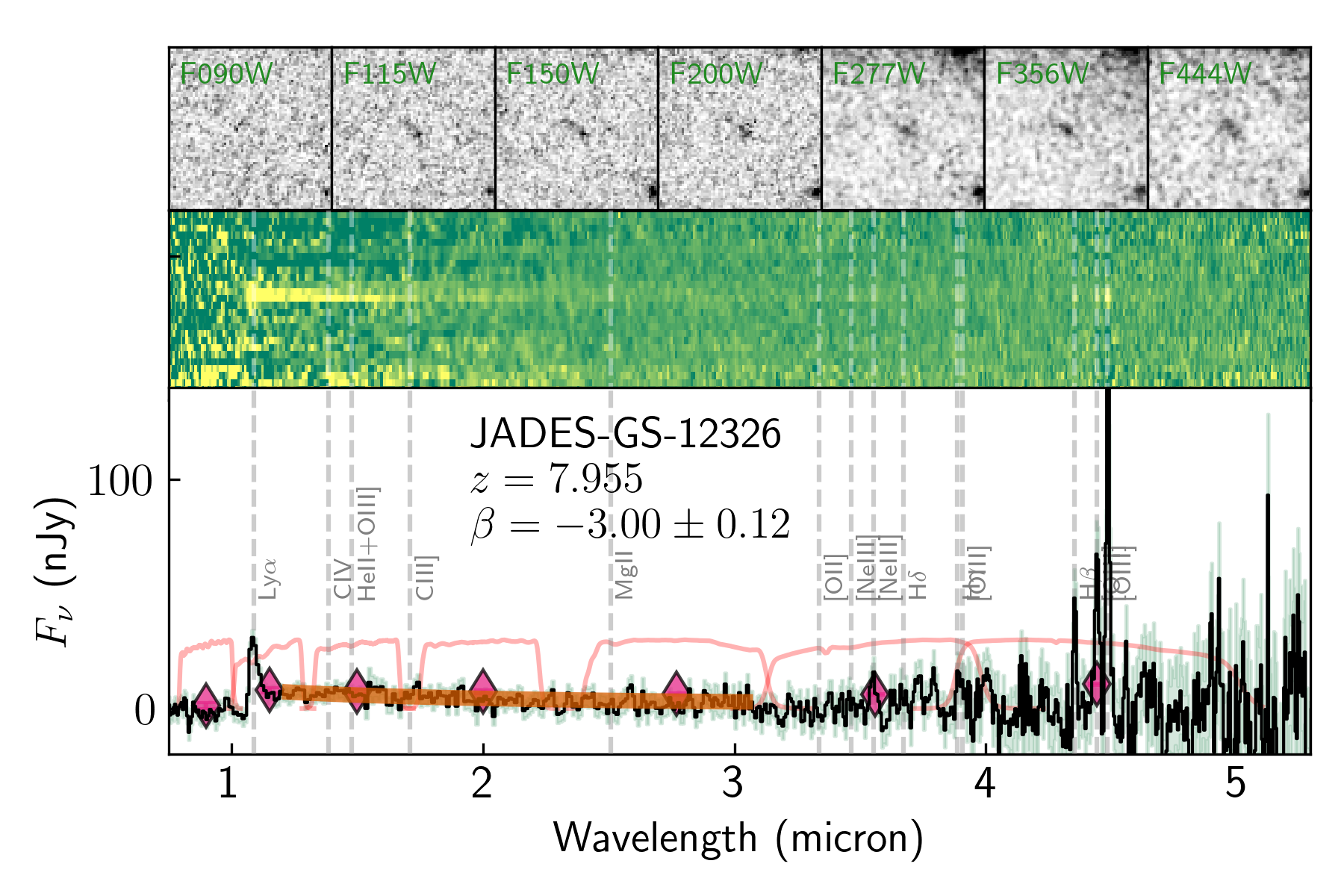}

\caption{1D and 2D spectra for the galaxies in our sample that show $\beta \leq -3.0$, selected from different tiers of JADES observations. Three out of these six galaxies show extremely bright \lya\ emission, which may be indicative of significant LyC escape as well \citep[e.g.][]{ver15}. GS-210003 at $z=5.779$ shows a very interesting spectrum, with clear detections of the Si\,\textsc{iv}+O\,\textsc{iv}] and \heii+\oiiiuv\ complexes. The Si\,\textsc{iv}+O\,\textsc{iv}] complex in particular is not widely detected across galaxy spectra, and if the O\,\textsc{iv}] is high, it would signify the presence of extremely hard ionizing radiation fields.}
    \label{fig:extreme-beta}
\end{figure*}

From our sample of 295 galaxies at $z>5.5$, we identify six galaxies that show extremely blue UV slopes with $\beta \leq -3.0$, which are likely tracing extremely young stellar populations, conditions that may enable high LyC escape fractions \citep[e.g.][]{top24a} and/or extremely low stellar/gas-phase metallicities \citep[e.g.][]{paa18}. In this section we look at the spectra of these six sources in more detail, measuring the relevant spectroscopic properties to determine what set of physical and chemical conditions might be responsible for the observed extremely blue UV slopes. We further estimate the LyC photon escape fractions (\fesc) using the $\beta$-based prescription derived by \citet{Chisholm2022} as well as the multi-variate \fesc\ estimation method derived using the \texttt{SPHINX} simulation by \citet{Choustikov2024a}. The \citet{Choustikov2024a} method takes as input the \hb\ luminosity, \muv, dust attenuation, O32 and R23 measurements, which are possible for our galaxies thanks to deep spectroscopy. We note that there are additional multi-variate estimators of LyC \fesc\ available in the literature \citep[e.g.][]{mas23, Jaskot2024b}, but given the available measurements for our sample we restrict ourselves to only the above mentioned two approaches.

The galaxies in this section are presented in decreasing blueness of their UV slopes, and their 1D (in $F_\nu$) and 2D spectra are shown in Figure \ref{fig:extreme-beta}, with the best-fitting $\beta$ slope also shown. For comparison, we further show the NIRCam photometry for these objects, which demonstrates how photometric bands crucial for measuring $\beta$ photometrically may often suffer from emission line or continuum feature contamination, which can impact the inferred $\beta$. 

\subsubsection{JADES-GS-16964}
This galaxy was observed as part of the GS-JWST-MEDIUM-1286 tier of JADES and with $\beta = -3.11 \pm 0.12$, JADES-GS-16964 has the bluest UV slope in our sample. This galaxy lies at a redshift of $z=7.19908$, and has $M_{\rm{UV}} = -19.21^{+0.35}_{-0.52}$. Strong \oiii\ emission is seen in the spectrum, but no clear \hb\ or \oii\ emission is detected, giving a limit on the O32 ratio of $>5.5$. Owing to a lack of strong emission lines from metals, this galaxy may be a strong candidate for an extremely metal poor system in the early Universe, with possibly significant LyC photon leakage suppressing the nebular continuum and line emission. Using the \citet{Chisholm2022} method we infer a very high LyC \fesc $=0.81\pm0.16$, whereas the multi-variate estimation using the \citet{Choustikov2024a} method gives a relatively lower but still significant LyC \fesc $=0.41\pm0.10$.

\subsubsection{JADES-GS-193797}
This galaxy was also observed as part of the GS-JWST-MEDIUM-1286 tier, and has $\beta = -3.08 \pm 0.10$. Several strong emission lines are visible in its spectrum, including the \ha, \oiii\ and \hb\ lines, giving a robust redshift of $z=5.84663$ and $M_{\rm{UV}} = -19.19^{+0.24}_{-0.30}$. As expected from the extremely blue UV slope, there is no dust attenuation measured from the Balmer decrement, and this galaxy has EW(\oiii\,$\lambda5007$) $>600\,$\AA, with O32 $>8.6$ and R23 $<14.6$. 

The \ha\ emission gives log(\xiion/Hz\,erg$^{-1}$) $= 25.3 \pm 0.2$. Using the \citet{Chisholm2022} method we estimate \fesc $=0.74\pm0.15$, with a comparable \fesc $=0.61\pm0.12$ estimated using the \citet{Choustikov2024a} method, which makes this galaxy a very likely LyC leaker.

\subsubsection{JADES-GS-210003}
Observed as part of the GS-3215 tier, this galaxy at $z=5.77885$ has $\beta=-3.02 \pm 0.03$, \muv\ $=-18.68^{+0.06}_{-0.06}$ and a strong \lya\ line with EW$_0$(\lya) $=75 \pm 10$\,\AA, comparable to the sample of LAEs that was presented by \citet{Saxena2024}. The much deeper exposure times in this tier result in high S/N detections of a host of rest-frame optical emission line features, including a very strong \oiii\,$\lambda5007$ line with EW(\oiii) $=3450 \pm 1170$\,\AA, O32 $=8.6 \pm 0.4$ and R23 $=7.8 \pm 0.3$. The strong \ha\ line has EW(\ha) $=2134 \pm 1171$\,\AA, giving an extremely high ionizing photon production efficiency, log(\xiion/Hz\,erg$^{-1}$) $=25.9 \pm 0.1$. There is no dust attenuation measured from the Balmer decrements.

We do note the presence of a possible contaminant in the 2D spectrum of this galaxy, which may have adversely affected the collapsed 1D spectrum from which the $\beta$ was measured. In the observed spectrum, a slight deviation from a pure power law can be seen at wavelengths just longer than 2 micron, which could either be due to contamination from the secondary source.

The spectrum also shows rest-frame UV features tracing high ionization conditions such as He\,\textsc{ii}+O\,\textsc{iii}] and C\,\textsc{iii}], as well as a strong Si\,\textsc{iv}+O\,\textsc{iv}] emission, which is seldom seen in galaxy spectra with such high S/N. If there is indeed a strong contribution from He\,\textsc{ii} and O\,\textsc{IV}] in the line complexes, this would signify the presence of extremely hard ionizing radiation sources, such as massive, hot stars, high-mass X-ray binaries, or even photoionization due to shocks \citep[e.g.][]{sax20, Lecroq2024}. The blue UV slope may help rule out photoionization from an AGN. Unfortunately, the \heii\ and \oiiiuv\ lines fall in the detector chip gap in the G140M spectra that were also obtained, and therefore the complex cannot be resolved with higher spectral resolution. The LyC \fesc\ derived from both our methods are highly comparable, with the \citet{Chisholm2022} method giving \fesc $=0.63\pm0.13$ and the \citet{Choustikov2024a} method giving \fesc $=0.60\pm0.12$.

\subsubsection{JADES-GS-3433}
This galaxy was observed as part of the GS-JWST-DEEP tier in JADES, and has $\beta = -3.02 \pm 0.06$ at $z=7.10868$, with \muv\ $=-18.60^{+0.16}_{-0.19}$.  This galaxy also shows \lya\ in emission with EW$_0$(\lya) $= 20 \pm 7.5$. The EW(\oiii) is $>930$\,\AA, and the \oii\ line detection gives O32 $=8.1 \pm 0.8$ and R23 $=6.5\pm0.7$. There is no dust attenuation inferred from the Balmer decrement. Using the \hb\ line we measure a relatively high log(\xiion/Hz\,erg$^{-1}$) $=25.8 \pm 0.1$. 

In the rest-frame UV, we see \ciii\ and \civ\ emission, and in the higher resolution G140M grating spectrum we also observe \heii, \oiiiuv\ and Si\,\textsc{iv}+O\,\textsc{iv}] emission lines, indicative of the presence of hard ionizing radiation fields. Once again, blue $\beta$ and a high O32 ratio could be indicative of LyC leakage. We also note that no dust attenuation was inferred using the Balmer emission lines. We once again measure high LyC \fesc\ values from both methods, with the \citet{Chisholm2022} method giving \fesc $=0.63\pm0.13$ and the \citet{Choustikov2024a} method giving \fesc = $0.54\pm0.11$.

\subsubsection{JADES-GS-116257}
This galaxy was observed as part of the GS-JWST-MEDIUM-1286 tier of JADES and has a $\beta = -3.02 \pm 0.11$ at $z=5.56377$ giving \muv $=-18.13^{+0.51}_{-0.98}$. The EW(\oiii) is $>1500$\,\AA, with an extremely high O32 ratio of $28.9 \pm 14.6$ and low R23 of $5.9 \pm 2.1$. We also measure EW(\ha) $> 1080$\,\AA, giving an extremely high log(\xiion/Hz\,erg$^{-1}$) $=26.0 \pm 0.2$. The [Ne\textsc{iii}] lines are also seen in this spectrum, consistent with the presence of hard ionizing radiation sources. This galaxy shows no dust attenuation from the Balmer decrement.

Owing to its relatively lower redshift, the \lya\ line falls in a noisier part of the spectral coverage of NIRSpec, and it therefore remains unclear whether there is strong \lya\ present. From the higher resolution G140M spectrum, we see hints of a possible \civ\ P-Cygni profile emission line, and weak \heii\ and \oiiiuv\ lines, although both of these are relatively uncertain. The high O32 ratio, high \xiion\ and blue $\beta$ point towards the presence of a spectrum dominated by light produced by extremely young (less than a few Myr) stars forming in a dust-free environment. 

Interestingly, the LyC \fesc\ estimates from the two methods used here differ significantly for this galaxy, with the \citet{Chisholm2022} method predicting \fesc $=0.63\pm0.13$ whereas the \citet{Choustikov2024a} method predicts a much lower \fesc = $0.12\pm0.04$. Given the extremely high EW of the nebular emission lines, and the extremely high \xiion, a lower LyC \fesc\ may be preferable such that a large fraction of ionizing photons are available to excite this strong line emission as opposed to leaking out of the galaxy into the IGM.

\subsubsection{JADES-GS-12326}
This source was observed as part of the deeper GS-JWST-DEEP tier of JADES, and shows a truly remarkable spectrum with $\beta = -3.00 \pm 0.12$ at $z=7.95477$, with a relatively faint \muv $=-17.71^{+0.10}_{-0.10}$. The PRISM spectrum shows an extremely strong and asymmetric \lya\ emission line with EW(\lya) $210 \pm 40$\,\AA, making it one of the highest redshift LAEs currently known \citep[e.g.][]{Saxena2024, Witstok2024}, which is likely tracing a large ionized bubble \citep[e.g.][]{sax23a, witstok23}.

This galaxy further shows strong \oiii\ and \hb\ lines, with EW(\oiii) $>1350$\,\AA, with no \oii\ emission seen in the spectrum, giving a limit of O32 $>15$ and R23 $<5.5$ suggestive of an extremely high ionization parameter and relatively low O/H abundance. From the \hb\ emission line assuming no dust, we measure an extremely high log(\xiion/Hz\,erg$^{-1}$) $=25.9 \pm 0.1$. No other rest-UV feature is clearly detected in the spectrum, which makes the presence of sources producing hard ionizing radiation fields harder to discern.

Once again, the two LyC \fesc\ predictors estimate very different values, with the \citet{Chisholm2022} method predicting \fesc $=0.59\pm0.12$ and the \citet{Choustikov2024a} method predicting a lower \fesc $=0.21\pm0.05$. The same arguments that were used for the previous galaxy could be applied here, where a lower \fesc\ may be preferable given the strong \oiii\ nebular line emission seen as well as the high \xiion\ inferred from the \hb\ emission. However, the presence of strong \lya\ emission may be indicative of a substantially higher \fesc. Unfortunately the \lya\ line in the higher resolution G140M falls in the detector chip gap, which makes it impossible to infer the \lya\ velocity offset compared to the systemic redshift, which is a good indicator of the LyC \fesc.

\subsubsection{Concluding remarks}
Overall, we find that galaxies with extremely blue $\beta \sim -3$ also on average show high O32 ratios, high \xiion, and presence of other high-ionization emission lines in their spectra, in line with expectations from young, massive stars being the dominant producers of ionizing photons in the galaxies. Four out of the six galaxies show strong \lya\ emission as well. The inferred LyC \fesc\ from the majority of these galaxies are high, in line with expectations that galaxies with blue UV slopes are good candidates for significant LyC leakage. In Table \ref{tab:steep_galaxies} we summarize the observed spectroscopic properties of these galaxies, and give the measured \xiion\ and LyC \fesc\ values as well.

\begin{table*}[]
    \centering
    \caption{Observed spectroscopic and derived properties of six galaxies in our sample that show extremely blue UV slopes ($\beta \leq -3.0$). We note here that none of these galaxies show any signs of dust attenuation from their Balmer decrement measurements. C22 \fesc(LyC) measurements are from the \citet{Chisholm2022} method, and the C24 measurements are derived from the \citet{Choustikov2024a} method.}
    \begin{tabular}{l c c c c c c c c c}
    \toprule
    ID & $z$ & $\beta$ & \muv & $F$(\hb) & O32 & R23 & log(\xiion) & \fesc(LyC) & \fesc(LyC) \\
     & & & & (\flux) & & & (Hz\,erg$^{-1}$) & (C22) & (C24) \\
    \midrule
    GS-16964 & 7.19908 & $-3.11 \pm 0.12$ & $-19.21^{+0.35}_{-0.52}$ & $-$ & $>5.5$ & $-$ & $-$ & $0.81\pm0.16$ & $0.41\pm0.10$ \\
    GS-193797 & 5.84663 & $-3.08 \pm 0.10$ & $-19.21^{+0.35}_{-0.52}$ & $3.4\times10^{-19}$ & $>8.6$ & $<14.6$ & $25.3 \pm 0.2$ & $0.74\pm0.15$ & $0.61\pm0.12$ \\
    GS-210003 & 5.77885 & $-3.02 \pm 0.03$ & $-18.68^{+0.06}_{-0.06}$ & $6.4\times10^{-19}$ & $8.6 \pm 0.4$ & $7.8 \pm 0.3$ & $25.9 \pm0.1$ & $0.63\pm0.13$ & $0.60\pm0.12$ \\
    GS-3433 & $7.10868$ & $-3.02 \pm 0.06$ & $-18.60^{+0.16}_{-0.19}$ & $3.0\times10^{-19}$ & $8.1 \pm 0.8$ & $6.5 \pm 0.7$ & $25.8 \pm 0.1$ & $0.63\pm0.13$ & $0.54\pm0.11$ \\
    GS-116257 & $5.56377$ & $-3.02 \pm 0.11$ & $ -18.13^{+0.51}_{-0.98}$ & $6.7\times10^{-19}$ & $28.9 \pm 14.6$ & $5.9 \pm 2.1$ & $26.0 \pm 0.2$ & $0.63\pm0.13$ & $0.12\pm0.04$ \\
    Gs-12326 & $7.95477$ & $-3.00 \pm 0.12$ & $-17.71^{+0.10}_{-0.10}$ & $1.2\times10^{-19}$ & $>15$ & $<5.5$ & $25.9 \pm 0.1$ & $0.59\pm0.12$ & $0.21\pm0.05$ \\
    \bottomrule
    \end{tabular}
    \label{tab:steep_galaxies}
\end{table*}

\section{UV slopes at redshifts $\mathbf{z\gtrsim9.5}$}
\label{sec:highz}
Having explored the connection between the UV slope and physical and chemical properties of galaxies at $z<9.5$ primarily traced by various spectroscopic indicators, in this section we turn our attention to the UV slopes of galaxies at $z>9.5$, where emission line detections still remain few and far between. 

\subsection{Insights from simple stellar population models}
\begin{figure*}
    \centering
    \includegraphics[width=\linewidth]{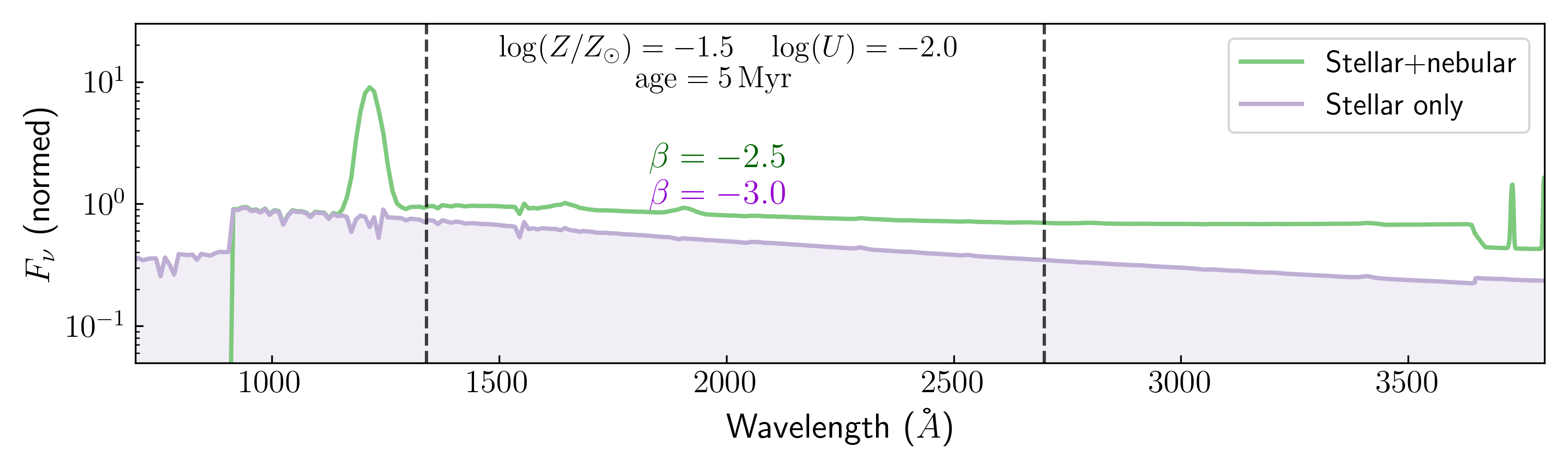}
    \caption{Synthetic spectra generated using \texttt{pythonFSPS} with both stellar and nebular emission (green) and stellar only emission (violet, shaded) for a simple stellar population with $\log(Z/Z_\odot) = -1.5$, $\log(U) = -2.0$ and age (time since burst) $= 5$\,Myr. The vertical dashed lines demarcate the wavelength range over which $\beta$ is measured. It is clear that the inclusion of the nebular continuum reddens the measured UV slope to $\beta = -2.5$ compared to a stellar only value of $\beta = -3.0$. The impact of reddening the $\beta$ due to nebular emission further depends on the physical and chemical conditions.}
    \label{fig:fsps_spectra}
\end{figure*}

Based on the statistical distribution of the measured UV slopes, it appears that the trend of increasingly bluer UV slopes as a function of redshift at $z<9.5$ appears to become less significant at $z>9.5$. Although there are selection effects at the highest redshifts that prevent the detection of UV-faint galaxies owing to the flux-limited nature of the spectroscopic surveys used here, the weak dependence of $\beta$ on the UV magnitude even in more `magnitude-complete' samples at lower redshifts in our sample suggests that incompleteness may not be the main cause of this reddening of UV slopes observed at the highest redshifts. The striking confirmation of a galaxy at $z=14.32$ presented in \citet{Carniani2024} showing a remarkably red UV slope of $\beta\approx-2.3$ presents a striking example. Therefore, in this section, we employ theoretical models to better understand the relatively red UV slopes seen across galaxies at $z>9.5$.

The first step in employing models to understand the observations is to produce a library of synthetic galaxy spectra that can then be directly compared to the observations. To do this, we use the \texttt{pythonFSPS} code\footnote{\url{https://dfm.io/python-fsps}} \citep{python_fsps}, which is a set of \texttt{python} bindings to the \texttt{FSPS} code\footnote{\url{https://github.com/cconroy20/fsps}} that allows the flexible modeling of stellar populations \citep{Conroy2009, Conroy2010}. Very briefly, \texttt{FSPS} allows the computation of simple stellar population (SSPs) for a wide variety of IMFs, metallicities, ages and ionization parameters, including the addition of nebular continuum \citep{Byler2017} and line emission. We use the standard installation of \texttt{FSPS} which uses the \texttt{MIST} isochrones \citep{Dotter2016, Choi2016} and \texttt{MILES} spectral library \citep{Sanchez2006, Falcon2011}. We further include the WM-Basic hot star library from \citet{eld17} for stellar temperatures above 25000\,K.

For simplicity, we assume a simple stellar population (SSP) model with a single burst, with the burst age extending to 40 Myrs, following a \citet{Chabrier2003} initial mass function (IMF) with an upper mass cutoff of $150\,M_\odot$. In all models, we include Wolf-Rayet star spectra as well. The SSPs are created over a metallicity range $\log(Z_\star/Z_\odot) = [-2.0, -0.5]$, and ionization parameters $\log(U) = [-3.0, -1.0]$. We also include trace amounts of dust with $E(B-V)$ values of $0.05$ and $0.10$, reflecting the level of dust attenuation seen in the $8<z<9.5$ bin from Balmer decrement measurements in Section \ref{sec: results}.

In Figure \ref{fig:fsps_spectra} we show example spectra with both stellar and nebular emission, and stellar only emission, for one of the models with $\log(Z/Z_\odot) = -1.5$, $\log(U) = -2.0$ and age (time since burst) $= 5$\,Myr. The vertical dashed lines indicate the wavelength range over which $\beta$ is typically measured. No dust attenuation has been added to these models. Clearly, the inclusion of nebular continuum emission can redden the measured UV slope from $\beta = -3.0$ for stellar only emission to $\beta = -2.5$ \citep[see also][]{Schaerer2005, Wilkins2013}. 

Having generated synthetic galaxy spectra for a variety of ages, metallicities and ionization parameters, we then measured the UV slope from these synthetic spectra in exactly the same way in which it was measured from observed spectra. As a result of this exercise, we have $\beta$ measurements for synthetic spectra containing only stellar emission as well as stellar plus nebular emission, over a range of galaxy physical properties, which can now be compared with our observations at $z>9.5$.

\subsection{The role of nebular continuum}

\begin{figure*}
    \centering
    \includegraphics[width=0.33\linewidth]{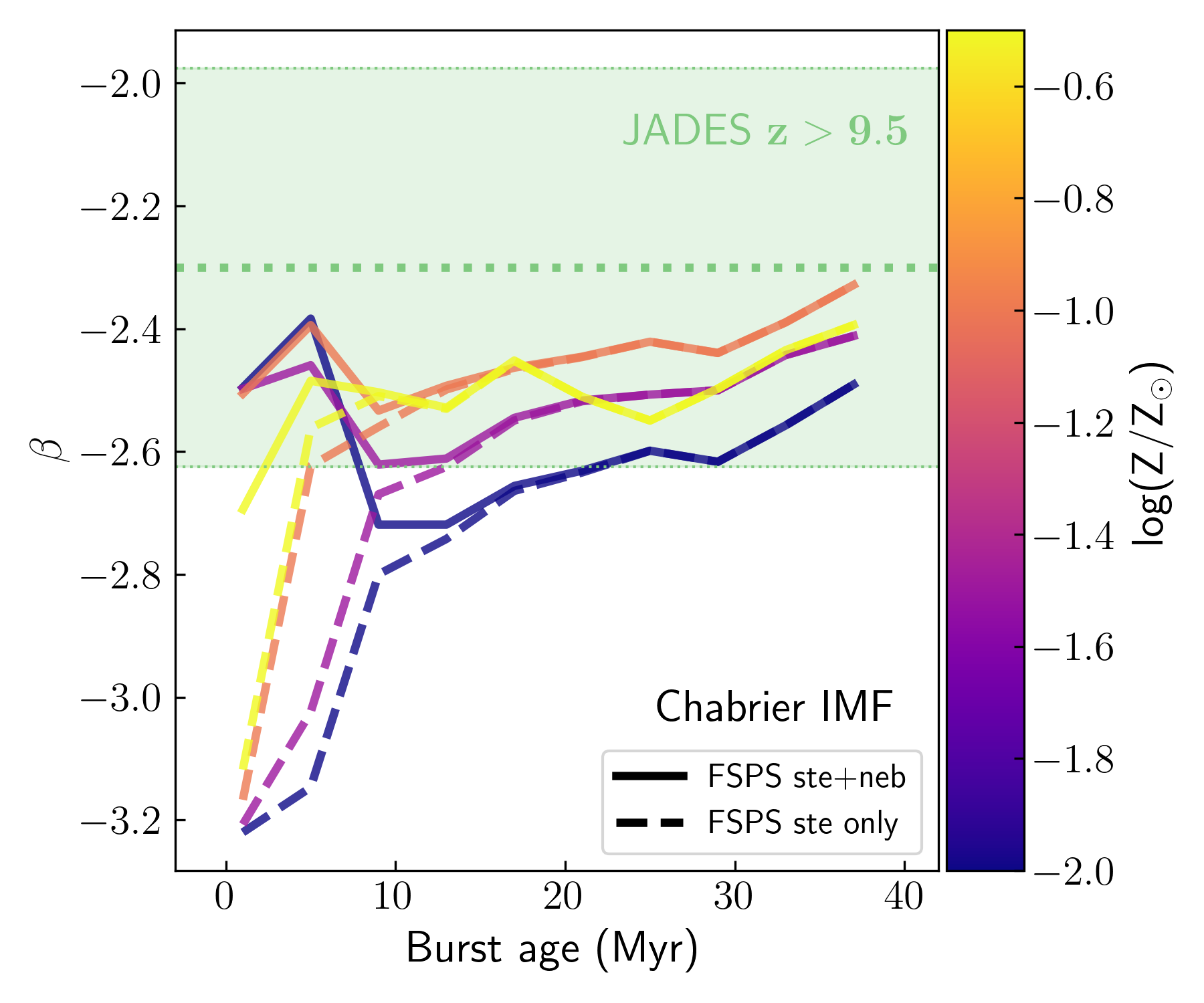}
    \includegraphics[width=0.33\linewidth]{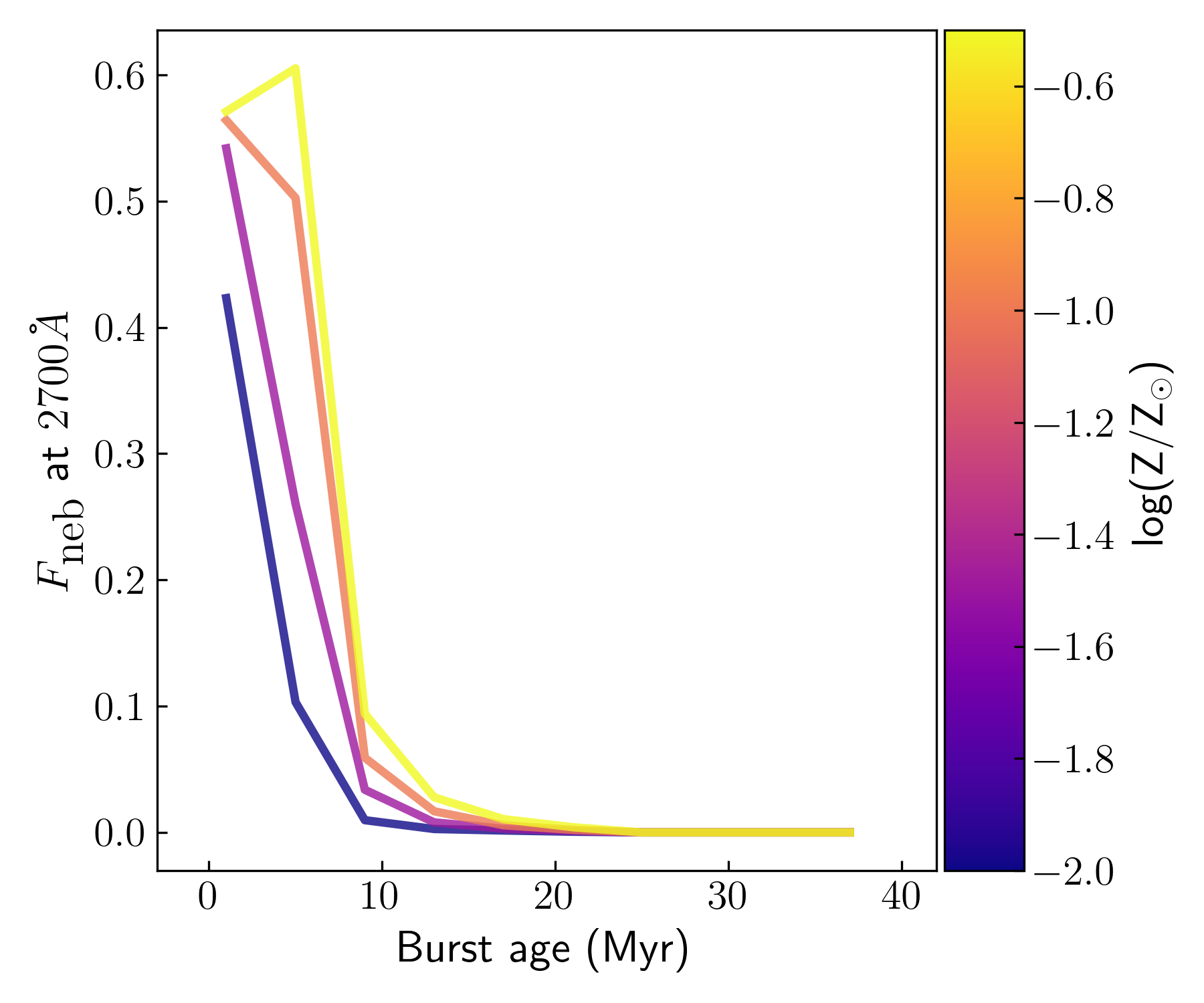}
    \includegraphics[width=0.33\linewidth]{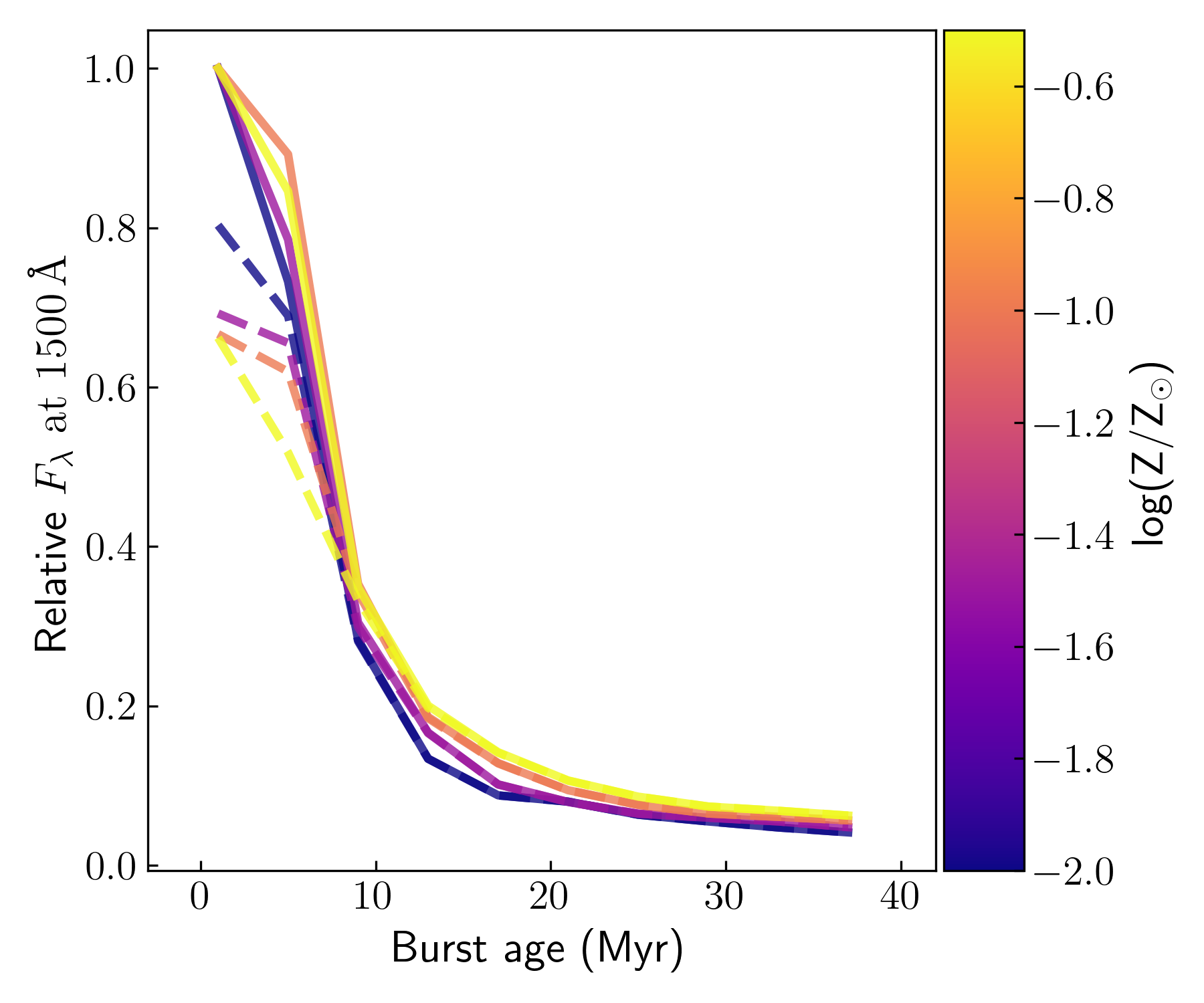}
    \caption{\emph{Left.} $\beta$ measured from synthetic spectra generated using \texttt{FSPS} with a Chabrier IMF as a function of age for a single burst of star formation with $\log(U) = -2.0$. Solid lines show measurements from spectra containing both stellar and nebular emission, and dashed line containing only stellar emission. The color bar represents metallicities in units of $\log(Z_\star/Z_\odot)$. The dotted line is the median observed $\beta$ for $z>9.5$ galaxies in our sample, and the shaded region represents the $1\sigma$ dispersion. \emph{Middle.} Fraction of nebular continuum measured at $2700$\,\AA, which leads to an overall reddening of the measured UV slope, as a function of burst age, color-coded by metallicity. We find that at low burst ages, nebular emission can contribute up to $60\%$ of the flux in the redder part of the wavelength range where $\beta$ is measured, thereby reddening the measured $\beta$ values. \emph{Right.} The relative $F_\lambda$ measured at $1500$\,\AA\ from the model predictions, normalized at the maximum flux recorded in the stellar+nebular models at zero age. The nebular continuum for this particular IMF can provide a boost of up to 50\% to the UV luminosity at young ages. Overall, we find that redder $\beta$ values can also be achieved at burst ages $>30$\,Myr, but this is unlikely to be the case at $z>9.5$ when galaxies are in the process of rapidly assembling their stellar masses through intense star-formation.}
    \label{fig:fsps-model}
\end{figure*}

We begin by investigating the impact of the nebular continuum on the observed UV slope of galaxies. A substantial fraction of nebular continuum can redden the UV slope of galaxies, even in the presence of young, hot stars whose black body emission peaks at shorter wavelengths, as was also recently demonstrated by \citet{Katz2024} for a sample of galaxies where the rest-frame UV spectrum appears to be dominated by the nebular continuum. In the upper panel of Figure \ref{fig:fsps-model}, we show the $\beta$ measured from synthetic spectra created following a Chabrier IMF as a function of time since burst for stellar only spectra (dashed lines) and spectra containing both stellar and nebular emission (solid lines) for $\log(U) = -2.0$ shown in the upper panel. The green dotted line marks the median $\beta$ measured from our $z>9.5$ galaxy sample, with the shaded region marking the $1\sigma$ deviation of the measurements.

It is clear that spectra that do not include any nebular contribution result in extremely blue UV slopes, particularly at early burst ages \citep[see also][]{cul23b}, which are incompatible with the median and $1\sigma$ dispersion of the measured $\beta$ at $z>9.5$ as shown in the left panel of Figure \ref{fig:fsps-model}. In the middle panel of Figure \ref{fig:fsps-model} we show the evolution of the fractional contribution of nebular emission at $2700$\,\AA, which generally represents the redder end of the wavelength range over which $\beta$ is measured, as a function of time since burst. We find that nebular continuum contributes up to $60\%$ of the flux at $2700$\,\AA\ (for this particular IMF and ionization parameter) at $\lesssim 10$\,Myr, thereby reddening the measured $\beta$. 

The $\beta$ from stellar only and stellar+nebular models converges at burst ages greater than $\sim10$\,Myr, when the nebular emission peters out. The reduction of the nebular continuum also results in a dramatic decrease in the UV luminosity at $1500$\,\AA\ for the same stellar mass formed, as shown in the right panel of Figure \ref{fig:fsps-model}. Therefore, to explain both the bright UV magnitudes and the red $\beta$ values seen in $z>9.5$ galaxies in this work and across the literature, either younger stellar ages or the addition of stellar mass at higher ages would be needed. The metallicity of the nebular gas plays an important role in further reddening the $\beta$, with lower metallicities of $\log(Z_\star/Z_\odot) < -1.0$ producing redder UV slopes at $<10$\,Myr since the burst. The trend is reversed at higher burst ages when the nebular emission is less prominent, and higher stellar metallicities produce redder $\beta$ values owing to cooler stellar temperatures.

Insights into relatively bursty star-formation histories of early galaxies both from theory \citep[e.g.][]{Faucher2018, Tacchella2020, Furlanetto2022, Sun2023, Mirocha2023} as well as observations of galaxies at the highest redshifts \citep[e.g.][]{end23b, top24a, Meyer2024} have highlighted the important role that young stellar populations play in shaping the continuum and emission line spectra of galaxies. Therefore, although higher stellar ages may be able to explain the relatively red $\beta$ values, owing to the star-formation histories of galaxies at $z>9.5$, increased fractional of nebular continuum driven by hot, massive forming stars may provide a more consistent explanation.

We find that none of the models can adequately explain the observed values redder than $\beta\gtrsim-2.3$, which is also what is measured for JADES-GS-z14-0 at $z=14.3$ \citep{Carniani2024}. From a stellar populations perspective, it may be possible to further redden the observed UV slope by including a considerably older stellar population in the synthetic spectra, however, at $z>9.5$ the Universe is less than 500\,Myr old, and the existence of an older stellar population already at these epochs may be in tension with expectations of first light in the Universe (typically expected to be $z\sim20-30$). Furthermore, no clear evidence of Balmer breaks in the spectra of galaxies at $z>9.5$ has yet been reported in the literature, casting doubt on the existence of older, more evolved stellar populations in these galaxies. 

The inclusion of dust attenuation of the rest-UV flux is another channel by which the observed $\beta$ values could be reddened. \citet{Carniani2024} showed that trace amounts of dust may be needed to produce $\beta = -2.3$ in JADES-GS-z14. Furthermore, \citet{Ferrara2024} recently proposed a scenario whereby galaxies can form a lot of stars very early in their lifetimes while being dust obscured leading to reddening of $\beta$, although this period lasts for a very short time before dust is eventually blown out by intense stellar feedback, making the UV slopes blue again. Therefore, in the following section we evaluate the role of dust attenuation in potentially reddening the UV slopes of $z>9.5$ galaxies. 

\subsection{The role of dust}

\begin{figure*}
    \centering
    \includegraphics[width=\linewidth]{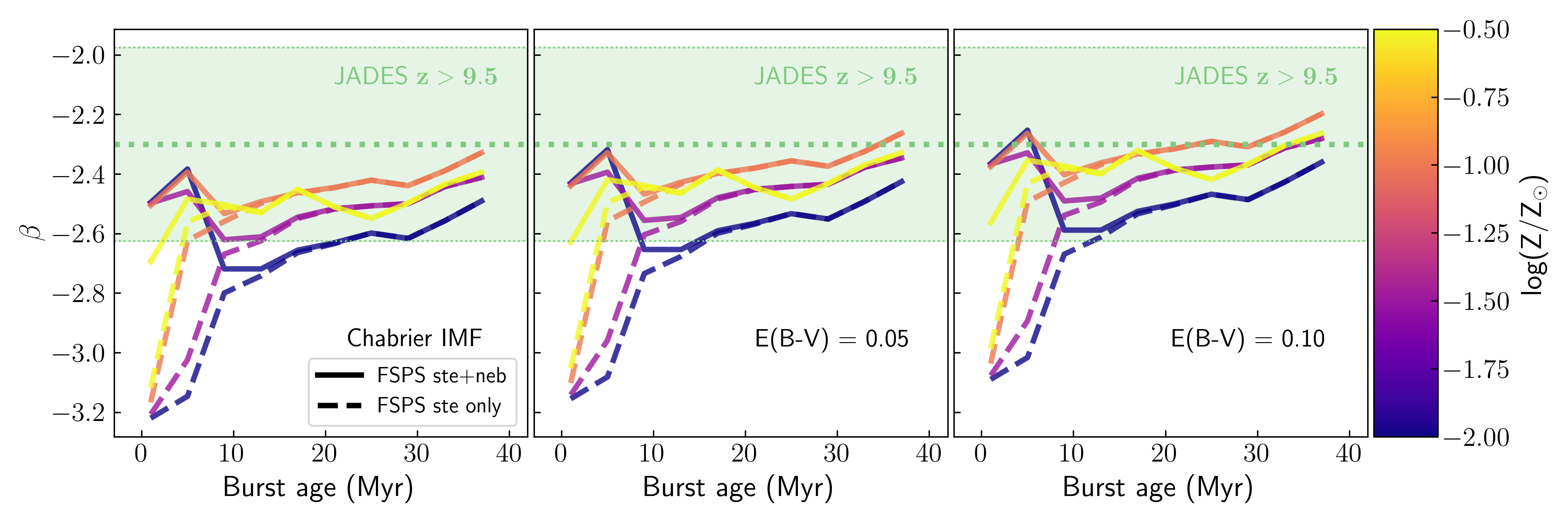}
    \caption{Same as Figure \ref{fig:fsps-model}, but here the different panels show the effects of dust attenuation (following the SMC attenuation law) on the UV slopes. The left panel shows the predictions from Chabrier IMF with no dust attenuation, the middle panel shows UV slope predictions with E(B-V) $=0.05$ and the right panel for E(B-V) $=0.10$. As expected, dust attenuation can redden the observed UV slopes, bringing the model predictions more in line with the measured UV slopes for our $z>9.5$ sample.}
    \label{fig:fsps-dust}
\end{figure*}

As shown in Figure \ref{fig:fsps-dust}, dust attenuation with $E(B-V)\sim0.1$ following the SMC extinction law can redden the UV slopes emerging from young stars, bringing them in line with the observed distribution seen in our sample. Modest dust attenuation with $E(B-V)\sim0.1$ was also seen from our measurements of the Balmer decrements in our $7<z<8$ subsample of galaxies, as shown in the top-left panel of Figure~\ref{fig:beta-physical}. 

In Figure \ref{fig:fsps-dust} we show the amount of reddening that $\beta$ may undergo with the introduction of dust attenuation with $E(B-V)=0.05$, corresponding to $A_V = 0.14$ following the SMC dust attenuation curve (middle panel) and $E(B-V) =0.10$ corresponding to $A_V = 0.28$ (right panel). We note that in our modeling, the stellar and nebular continuum spectra are affected by dust attenuation in the same manner. Clearly, the observed $\beta$ values at $z>9.5$ require more significant dust attenuation, which was typically seen across our $5.5<z<8$ galaxies inferred from the Balmer decrements.

However, the physical arguments behind the importance of the role of dust specifically at $z>9.5$ are more nuanced. The main challenge with achieving dust attenuation at these early redshifts is primarily dependent on rapid dust production through stellar evolution \citep[e.g.][]{Valiante2009, Gall2011}, growth, and the prevention of dust destruction from intense stellar feedback. Core-collapse supernovae are contemplated to be the most likely sources of dust production at early epochs, due to their short lifetimes and production of metals \cite[e.g.][]{Nozawa2010}. Additionally, intermediate and high-mass AGB stars (with masses between $3-8\,M_\odot$) that have sufficiently short lifetimes ($10^7-10^8$\,yr) may also be potential contributors to dust production \citep[e.g.][]{Schneider2014}, although their contribution is highly dependent on the formation timescales of the first stars in galaxies.

\citet{Narayanan2024} recently showed that trace amount of dust attenuation may indeed be reflected in the UV slopes of galaxies at $z>9.5$, however the fraction of galaxies that show an appreciable dust reddening rapidly plummets at redshifts higher than 9.5. Beyond $z>10$, dust masses $>10^6\,M_\odot$ were not seen in the simulations of \citet{Narayanan2024}, which resulted in minimal dust reddening of the UV slopes.  Dust attenuation certainly becomes more pronounced at lower redshifts due to enhanced grain growth per unit dust mass out to redshifts of $\sim6$. Therefore, it remains unclear whether dust attenuation represents the dominant reddening mechanism for UV slopes of galaxies particularly at $z>9.5$. As \citet{Narayanan2024} note, nebular continuum can efficiently redden the UV slope by an additional $\Delta\beta = 0.2-0.4$, whereas dust attenuation at $z>9.5$ only leads to $\Delta\beta < 0.1$. Owing to challenges with the rapid production, growth and preservation of dust in very high redshift galaxies, where it may be easy to destroy and displace dust due to stellar feedback \citep[e.g.][]{Ferrara2024}, high dust masses leading to significant reddening of the observed UV slopes might not be feasible.

\subsection{The role of the initial mass function (IMF)}

\begin{figure*}
    \centering
    \includegraphics[width=\linewidth]{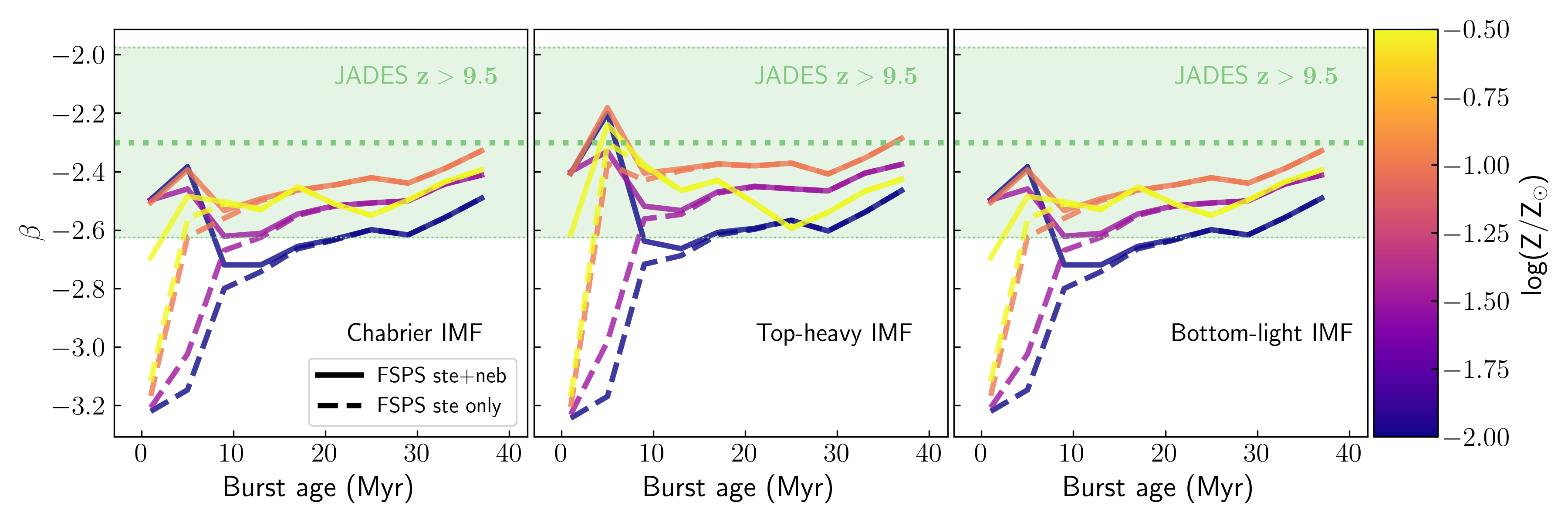}
    \caption{Same as Figure \ref{fig:fsps-model}, with the left panel showing predictions using Chabrier IMF, middle panel showing top-heavy IMF and the right panel showing outputs from a bottom-light IMF. The parameterization of these non-standard IMFs can be found in the text. Qualitatively speaking, there is not much difference between the predictions from Chabrier and bottom-light IMF, but a top-heavy IMF is able to produce redder $\beta$ values for younger burst ages, getting closer to the median $\beta$ measured for our $z>9.5$ galaxies. Redder $\beta$ values are likely driven by the presence of a larger number of hot, massive, but short-lived stars in a top-heavy IMF.}
    \label{fig:fsps-imf}
\end{figure*}

In this section, we explore the role of the IMF in setting the observed UV slopes, in the absence of substantial dust attenuation. In addition to the standard \citet{Chabrier2003} IMF with an upper mass cutoff of $150\,M_\odot$ that was employed earlier, we introduce two new IMF prescriptions here. The first `non-standard' IMF that is the so-called `top-heavy' IMF, with a lower mass ($M_\star < 0.8\,M_\odot$) slope of $-1.3$ and a high mass slope of $-1.6$ \citep[e.g.][]{Dabringhausen2009} with a maximum stellar mass of $600\,M_\odot$, where the slope describes the distribution function $dn \propto m^\gamma dm$, where $m$ is the birth mass of the star \citep[e.g.][]{Fukushima2023}. It has been argued that the `turn-over' of the IMF is dependent on the minimum temperature of the molecular clouds out of which stars form, which in turn is linked to the cosmic microwave background temperature, $T_{\rm{CMB}}$ \citep[e.g.][]{Bastian2010}. The increasing $T_{\rm{CMB}}$ at high redshifts will, therefore, result in lesser fragmentation in star-forming clouds, potentially leading to larger number of high mass stars and a more top-heavy IMF \citep[e.g.][]{Klessen2023}.

We additionally include a `bottom-light' IMF, which instead of boosting the high-mass end of the IMF compared to a standard Chabrier/Salpeter IMF results in suppressing the lower mass end of the IMF. Evidence for a bottom-light IMF in massive star clusters has been recently presented by \citet{Baumgardt2023}, and we use their parameterization here. For low mass stars with $M_\star < 0.4\,M_\odot$, the slope is $-0.3$, for intermediate mass stars in the mass range $0.4 < M_\star/M_\odot < 1.0$ the slope is $-1.65$, and for high mass stars with $M_\star > 1.0\,M_\odot$, the slope of the IMF is $-2.3$.

In Figure \ref{fig:fsps-imf} we show the evolution of $\beta$ as a function of burst age at $\log(U) = -2.0$, with the standard Chabrier IMF output on the left, the top-heavy IMF in the middle and the bottom-light IMF on the right. It is clear that qualitatively speaking, there is no appreciable difference between the Chabrier and bottom-light IMFs. However, the top-heavy IMF implementation results in overall redder $\beta$ slopes across all burst ages, even achieving $\beta=-2.2$ that no other model could achieve. The reddest $\beta$ slopes are once again only obtained via the inclusion of nebular continuum at ages younger than 10\,Myr. The effect of metallicity across all IMFs is broadly consistent with one another, with extremely low metallicities capable of producing redder $\beta$ slopes at low ages with the inclusion of nebular continuum. However at later times, once the fractional contribution of nebular emission is next to negligible, lower metallicities result in bluer $\beta$ values.

This exercise demonstrates that a top-heavy IMF does help bring the model-predicted $\beta$ closer to that which is observed from galaxies at $z>9.5$ by significantly boosting the fractional contribution of the nebular continuum to the rest-frame UV spectra. However, although a larger number of hotter, high mass stars are produced in a top heavy IMF, they do not live for long enough \citep{Klessen2023} to sufficiently heat up the nebular gas and redden $\beta$ for longer periods of time after the initial burst. Therefore, to both redden the $\beta$ as well as maintain a high enough total UV luminosity, inclusion of significant nebular continuum at low burst ages may be preferred explanation \citep[see also][]{Katz2024}.

\subsection{The role of gas temperature and density on the nebular continuum}
\begin{figure}
    \centering
    \includegraphics[width=\linewidth]{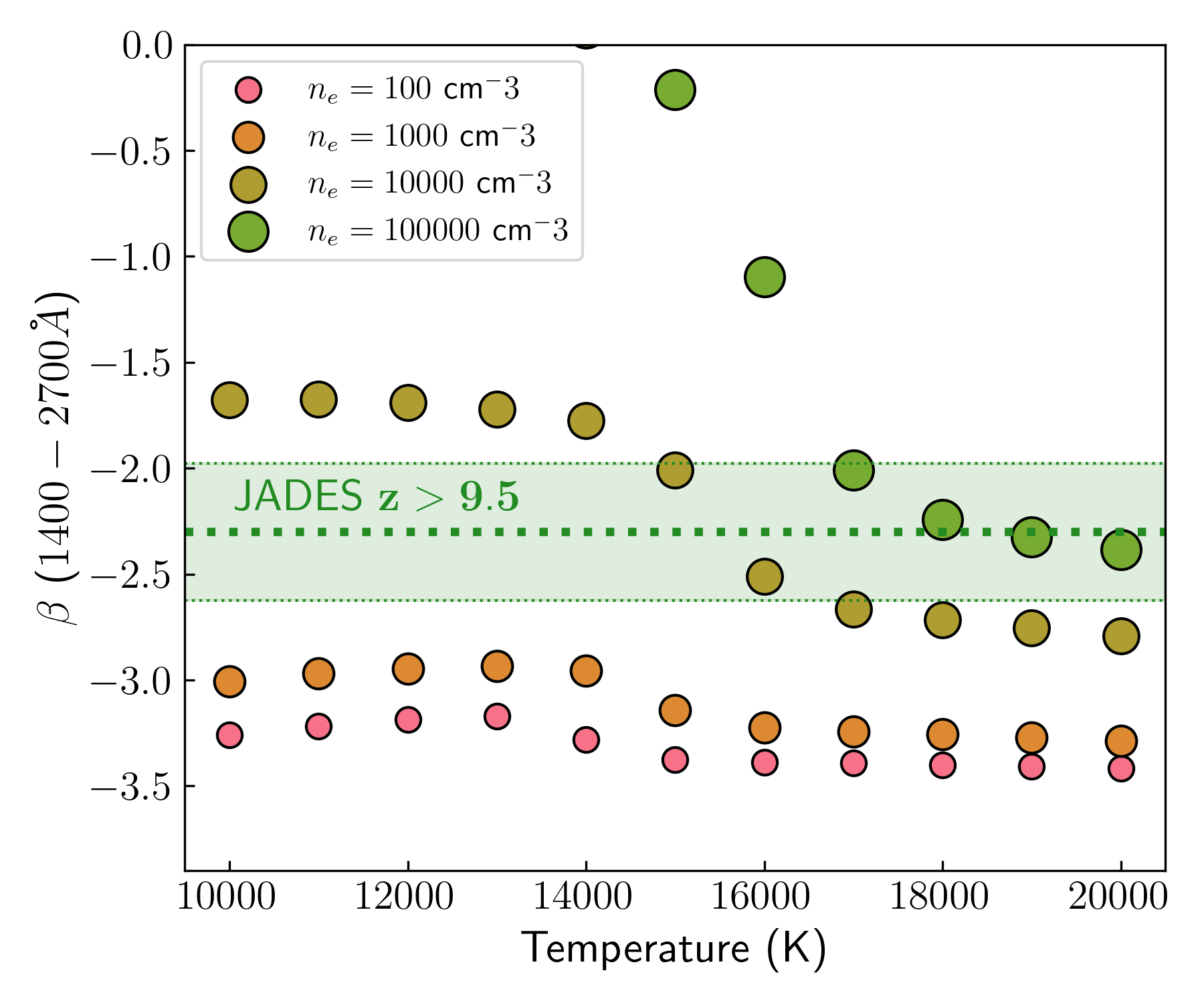}
    \caption{Dependence of $\beta$ with gas temperature for a purely nebular dominated rest-frame UV spectrum over the wavelength range where $\beta$ is typically measured, for a range of electron densities. We find that to reproduce the observed distribution of $\beta$ values seen at $z>9.5$ from our sample in the typically observed density range $n_e = 100-1000$\,cm$^{-3}$, a purely nebular dominated UV continuum from gas at temperatures above $15000$\,K is needed. In reality, however, the observed spectrum will contain only a fractional contribution of nebular continuum, but higher temperatures are nonetheless needed to redden the $\beta$ values.}
    \label{fig:nebular-predictions}
\end{figure}

Building on the finding that the nebular continuum component may be important to redden the UV slopes of galaxies at $z>9.5$, in this section we qualitatively explore how the strength and shape of the nebular continuum may vary with changes in the physical properties, particularly the temperature and density, of the nebular gas, which in turn will be driven by stars with different masses and temperatures. \textit{JWST} observations have shown that the ISM density increases dramatically from $z\sim0$ to $z\sim9$ \citep{Isobe2023}. 

Additionally, it has also been shown that the ionizing photon production rates in galaxies increase steadily towards high redshifts \citep[e.g.][]{sim24, Simmonds24b}, implying that the underlying stellar populations likely contain hotter, more massive stars that may also be capable of heating up the gas to higher temperatures than those that might be seen at lower redshifts, resulting in a greater fractional contribution from the nebular continuum to the overall UV spectrum \citep[e.g.][]{kat23b, cur23, san23, Laseter2024, Cameron2024_Neb}. Additional heating sources such as high-mass X-ray binaries, which become more important at lower metallicities at high redshifts \citep[e.g.][]{sax21}, or even the presence of AGN may further aid in heating the gas to higher temperatures.

\citet{Katz2024} recently reported a sample of spectroscopically confirmed star-forming galaxies that show signatures of a nebular-dominated continuum. Using SPS and photoionization models, \citet{Katz2024} showed that the fractional contribution of the nebular continuum to the observed rest-UV continuum of galaxies critically depends on stellar temperatures, as well as the temperature and density of the gas. As we previously demonstrated, the IMF can also play a crucial role in boosting the nebular continuum.

To explore how the nebular continuum may change under different physical conditions, in this section, we adopt a simpler approach compared to \citet{Katz2024} and create synthetic nebular continuum and line emission spectra using the \texttt{NEBULAR} code\footnote{\url{https://github.com/mischaschirmer/nebular/}} \citep{Schirmer2016}. \texttt{NEBULAR} allows the synthesis of nebular continuum and line emission from a mixed Hydrogen and Helium gas in collisional ionization equilibrium. \texttt{NEBULAR} can create spectra over a range of temperatures and densities, and includes contribution from free-free, free-bound, two-photon and line emission from H\,\textsc{i}, He\,\textsc{i} and \heii. The goal of this exercise is simply to qualitatively investigate what kinds of physical conditions in the ISM can preferentially redden the observed UV slopes, under the assumption that the observed galaxy spectrum may contain a considerable fraction of nebular continuum emission.

In Figure \ref{fig:nebular-predictions} we show the measured $\beta$ from synthetic nebular only spectra generated using \texttt{NEBULAR}, over an electron density range of $n_e = [100, 100000]$\,cm$^{-3}$ (in log space) and temperature range $T = [10000, 20000]$\,K. The size of the marker scales with density, with the smallest point having the lowest density and the largest point having the highest density. 

The nebular emission from gas below temperatures of $T<15000$\,K produces a very blue spectrum for densities $<10000$\,cm$^{-3}$, and no amount of fractional contribution from such nebular emission will significantly redden the observed UV slope of galaxies. For typical densities of $n_e = 100-1000$\,cm$^{-3}$ that are measured from spectra of high redshift galaxies, the nebular continuum emission will redden the observed galaxy spectrum only at gas temperatures of $T>15000$\,K, capable of explaining the range of $\beta$ values seen in our $z>9.5$ spectroscopic sample via a purely nebular dominated UV continuum.

A higher density can also lead to redder UV slopes at any given temperature, as higher densities significantly boost the free-bound and the free-free contribution to the nebular continuum. We note here that over the wavelength range that is used to measure the UV slope, the two-photon component does not have an appreciable impact. As also discussed by \citet{Cameron2023b}, the two-photon component of the nebular continuum is more prominent at lower electron densities ($n_e < 1000$\,cm$^{-3}$), but at higher densities the reddening of $\beta$ is likely caused by the free-bound component.

\subsection{Implications for the observed galaxy properties at $z>9.5$}
Thus far we have demonstrated that the UV slopes of spectroscopically confirmed galaxies at $z>9.5$ appear to redden, deviating from the trend of bluer $\beta$ observed with increasing redshift observed at $z<9.5$. We have also shown that to explain these relatively redder UV slopes in some of the highest redshift galaxies, inclusion of nebular continuum emission and/or dust attenuation may be required, particularly when stars forming out of relatively metal free gas would be expected to be more massive and hotter, with the stellar emission peaking at even shorter wavelengths, leading to a blue $\beta$ from starlight alone. Trace amounts of dust ($A_V ~ 0.2-0.3$) may be able to explain the observed distribution of $\beta$ at $z>9.5$, but producing and growing enough dust at these early epochs is challenging from a modeling perspective \citep[see][for example]{Narayanan2024}.

Furthermore, we have demonstrated that with increasing gas temperatures and densities, the nebular continuum emission can take on redder UV slopes. For the median $\beta = -2.3$ observed for our $z>9.5$ galaxy sample, pure nebular emission at $T>15000\,$K is able to explain both the median and the $1\sigma$ scatter in our $\beta$ measurements for this sample. This suggests that, in the absence of appreciable amounts of dust, bright nebular continuum from hot and dense gas driven by considerably hotter stars can redden the observed UV slopes of galaxies, particularly when the fractional contribution of the nebular continuum is high. Indeed a few examples of nebular dominated galaxy spectra across different redshifts have been previously presented by \citet{Fosbury2003, Cameron2024_Neb, Mowla2024, Katz2024}, where little to no stellar emission is needed to explain the observed UV continuum.

Naturally an increased fractional contribution of nebular emission to the continuum light of a galaxy, particularly at high redshifts when the gas temperatures are high, will have implications on the inferred galaxy properties from observations. We begin our discussion by tackling the question of the extremely bright absolute UV magnitudes that have been observed at record distances, particularly in sources such as GN-z11 \citep{bun23}, GS-z12 \citep{Deugenio2023} and GS-z14-0 \citep{Carniani2024} in our sample, as well as GHZ2 \citep{Castellano2024}. Interestingly, all of these galaxies show relatively red $\beta$ slopes ($-2.0$ to $-2.5$). A further overabundance of photometrically-selected UV-bright galaxies at $z>10$ has also been reported widely \citep[e.g.][]{Robertson2023, Donnan2024, Mcleod2024, Adams2024, Chemerynska2024}. If the UV continuum of these galaxies does contain a higher fraction of nebular emission \citep[e.g.][]{Katz2024}, then the observed UV magnitudes will be brighter without the need to boost the underlying stellar masses via extremely high star-formation efficiencies \citep[e.g.][]{Inayoshi2022, Dekel2023}. Such a scenario may explain both the extremely bright UV magnitudes of a large number of $z>10$ galaxies, as well as their relatively redder UV slopes.

With the exception of GN-z11, GS-z12 and GHZ2, the vast majority of spectroscopically confirmed galaxies at $z>10$ do not show strong rest-frame UV emission lines in their spectra \citep[e.g.][]{Finkelstein2022, curtis23, Arrabal2023, Hainline2024, Carniani2024}. If the UV continuum is boosted by nebular emission at higher gas temperatures, then this may play a role in lowering the observed EWs of the rest-frame UV emission lines, making them harder to detect even with deep \emph{JWST} spectroscopy. Other plausible explanations for the non-detection of emission lines in the vast majority of galaxies are either extremely low metallicities or high LyC escape fractions. A high LyC escape fraction will make the UV slope of the galaxies bluer than what is typically observed by suppressing the contribution of both the nebular continuum and line emission from the spectrum \citep[e.g.][]{zac13}. Lower metallicities may therefore be preferable as an explanation, which as we showed in Figure \ref{fig:fsps-model} will also self consistently lead to redder $\beta$ values at lower burst ages.

We showed that by increasing the gas temperatures in the ISM, which can theoretically be probed by observing the auroral line transitions of some of the most common species, overall redder UV slopes from nebular continuum emission can be achieved. Therefore, to heat the gas to such high temperatures, hotter ionizing sources such as massive Wolf-Rayet stars \citep[e.g.][]{Rivera2024} that can reach temperatures of up to $200000$\,K may be present. Very massive stars (VMS; $M_\star > 100\,M_\odot$) can also lead to both high effective temperatures and luminosities \citep{Sabhahit2022}, with their higher mass loss rates and strong winds also leading to rapid enrichment of the gas \citep[e.g.][]{Vink2011}. Further non-thermal heating, such as that provided by high-mass X-ray binaries that become dominant at low metallicities and higher binary fractions \citep{sax21, Umeda2022, kat23b}, may also become important \citep[see also][]{Katz2024}.

Finally, we also note again that for our $z<9.5$ sample, based on the relationship between O32 and $\beta$ we inferred that more complicated ISM geometries may need to be considered \citep[e.g.][]{Katz2022Prism, Jin2022} or prevalence of ionization-bounded \hii\ regions may be needed to explain the decrease in O32 ratios at the bluest UV slopes. These scenarios become important when the galaxies are undergoing periods of intense star-formation that introduces turbulence in the ISM. Galaxies at $z>9.5$, owing to their youthfulness, will likely also experience highly turbulent ISM conditions, and it remains unclear what impact these effects have on the ISM of $z>9.5$ galaxies, and what it means for the predicted nebular continuum emission and effect on observed UV slopes.

\section{Summary}
\label{sec:summary}
In this study, we have presented spectroscopic measurements of the rest-frame UV slope, $\beta$, for a sample of 295 galaxies at $z>5.5$ using \emph{JWST}/NIRSpec observations, using both individual measurements as well as measurements from stacked spectra created in bins of redshift and $\beta$. All our galaxies have robust spectroscopic redshifts, which enables a highly reliable estimate of $\beta$ for our sample. The main aims of this study are to characterize which physical and chemical conditions are responsible for setting the observed $\beta$ in high redshift galaxies, and what the global evolution of $\beta$ as a function of galaxy properties and redshifts can tell us about the evolving properties of galaxies over cosmic time.

Across our sample, we find a median $\beta$ value of $-2.30$ and a mild increase in the blueness of $\beta$ with redshift. On the other hand, we find evidence of increased reddening of $\beta$ at $z>9.5$ compared to its evolution at lower redshifts. We further find a weak trend of increasingly bluer $\beta$ at fainter absolute UV magnitudes. This trend between $\beta$ and UV magnitude is found to be stronger at $z<8$, with the trend breaking down at $z>8$. Overall we find that the range of $\beta$ values remains large and consistent across redshifts.

Based on spectroscopic measurements performed on spectra stacked in bins of $\beta$ and redshift, we find that galaxies with the bluest $\beta$ generally have little to no dust inferred from the Balmer decrements, with nearly all galaxies above $z>8$ exhibiting E(B-V) $<0.1$ regardless of their $\beta$ slope. The UV slope does not show any clear trend with gas-phase metallicities. When comparing the \oiii\,$\lambda5007$/\oii\,$\lambda3727$ (O32) ratios, we find that galaxies with bluer $\beta$ generally show higher O32 ratios. However, for galaxies between redshifts of $8<z<9.5$, we find that the O32 ratio decreases for the bin with the bluest $\beta$ ($\beta\approx -2.6$) compared to what is seen for the intermediate $\beta$ bin ($\beta \approx -2.2$). Finally, we do not see any clear links between $\beta$ and the strength of the Balmer emission lines across our bins.

From our sample, we identify six galaxies that have ultra-blue $\beta<-3.0$, which all show interesting spectroscopic features. These galaxies span a redshift range of $z=5.5$ to $z=8.0$, and three out of the six galaxies show strong \lya\ emission, including the highest redshift galaxy from these six at $z=7.996$. All but one of these six galaxies show strong rest-frame optical nebular emission lines and high O32 ratios, which favors the existence of density-bounded \hii\ regions with possibly signifincant LyC photon leakage. The galaxy with the bluest slope of $\beta=-3.11$ shows very weak rest-optical lines, which may favor an extremely low metallicity and/or high LyC escape fraction.

We then explore $\beta$ values in our $z>9.5$ galaxy subsample, finding a median $\beta=-2.3$, with $1\sigma$ dispersion ranging from $\beta-2.0$ to $\beta=-2.6$. When comparing with simple stellar population models, we find that models that only take into account emission from young, metal-poor stars produce $\beta$ slopes that are too blue when compared with the observed values. Therefore, inclusion of nebular continuum is necessary to get closer to explaining the observed $\beta$ values at $z>9.5$, but some of the reddest observed $\beta$ values still remain out of reach of these models. We show that inclusion of a trace amount of dust attenuation can help redden the model $\beta$ slopes. Furthermore, in the absence of dust attenuation, using a more top-heavy initial mass function also results in overall redder UV slopes.

Focusing on the role of the nebular continuum in reddening the UV slopes, we then show that increasing the gas temperatures above $15000$\,K over a density range $n_e = 100-1000$\,cm$^{-3}$ can reproduce the entire $1\sigma$ range of the observed $\beta$ values for our $z>9.5$ sample, assuming that the observed UV continuum is dominated by the nebular continuum. We find that achieving such high gas temperatures within these galaxies may require additional sources of heating, such as massive Wolf-Rayet stars, very massive stars ($M_\star > 100\,M_\odot$) or non-thermal heating sources such as high-mass X-ray binaries or AGN. Inclusion of these additional populations of sources in stellar population synthesis models may be needed to self consistently explain the observed galaxy properties and the $\beta$ slope measurements at the highest redshifts.

In this work, we have demonstrated that a purely spectroscopic analysis of the UV slope of a large sample of galaxies at $z>5.5$ is now possible thanks to a number of large spectroscopic surveys that \emph{JWST} has been carrying out over the last few years. Establishing the conditions that drive the all-important rest-UV slope of galaxies in the presence of valuable spectroscopic information for galaxies at the highest redshifts is a remarkable step forward for high-redshift galaxy evolution studies, and more detailed modeling may be needed to understand the spectroscopic and photometric properties of some of the earliest galaxies that formed in the Universe.

\section*{Acknowledgements}
AS would like to thank Nick Choustikov for useful discussions about LyC leakage from galaxies with blue UV slopes. AS thanks Richard Ellis for fruitful discussions. AS, AJC, AJB and JC acknowledge funding from the “FirstGalaxies” Advanced Grant from the European Research Council (ERC) under the European Union’s Horizon 2020 research and innovation programme (Grant agreement No. 789056). FDE, GCJ, CS, and ST acknowledge support by the Science and Technology Facilities Council (STFC) and by the ERC through Advanced Grant number 695671 ``QUENCH'', and by the UKRI Frontier Research grant RISEandFALL. SA acknowledges support from the research project PID2021-127718NB- I00 of the Spanish Ministry of Science and Innovation/State Agency of Research (MICIN/AEI). SC acknowledges support by European Union's HE ERC Starting Grant No. 101040227 - ``WINGS''. ECL acknowledges support of an STFC Webb Fellowship (ST/W001438/1). BDJ and BER acknowledge support from the NIRCam Science Team contract to the University of Arizona, NAS5-02015. H\"U acknowledges support through the ERC Starting Grant 101164796 ``APEX''. The work of CCW is supported by NOIR-Lab, which is managed by the Association of Universities for Research in Astronomy (AURA) under a cooperative agreement with the National Science Foundation.  

This work is based on observations made with the NASA/ESA/CSA James Webb Space Telescope. The data were obtained from the Mikulski Archive for Space Telescopes at the Space Telescope Science Institute, which is operated by the Association of Universities for Research in Astronomy, Inc., under NASA contract NAS 5-03127 for JWST. These observations are associated with program \#1180, \#1181, \#1210, \#1286, \#1287 and \#3215.

This work has made extensive use of the \texttt{astropy} \citep{astropy, astropy2}, \texttt{matplotlib} \citep{plt} and \texttt{pandas} \citep{pandas} \texttt{python} packages. This work would not have been possible without the hard work of thousands of developers and volunteers around the world, who foster a culture of open source software development.

\section*{Data Availability}

The high-level science products (HLSP) underlying this analysis are either available in full or in part both from the JADES collaboration webpage (\url{https://jades-survey.github.io}) and from the Mikulski Archive for Space Telescope (\url{https://archive.stsci.edu/hlsp/jades}). HLSP that are currently not publicly available will be made so at the end of the 12-month proprietary period, as per the GTO agreements. The data analysis pipelines will shortly be made available on the lead author's GitHub page (\url{https://github.com/aayush3009}).



\bibliographystyle{mnras}
\bibliography{jades} 




\appendix

\section{Stacked 1D rest-frame spectra of galaxies in redshift-$\beta$ bins}
Here we show the stacked rest-frame spectra in bins of redshift and $\beta$, as described in Table \ref{tab:stack_bins}. The fluxes are normalized in units of $F_\nu$. The stacked spectra are represented by the solid green lines in each bin, with the individual (normalized) spectra going in to the stack shown as translucent gray lines in the background. Also marked are the positions of strong/common emission line features.
\begin{figure*}
    \centering
    \includegraphics[width=\linewidth]{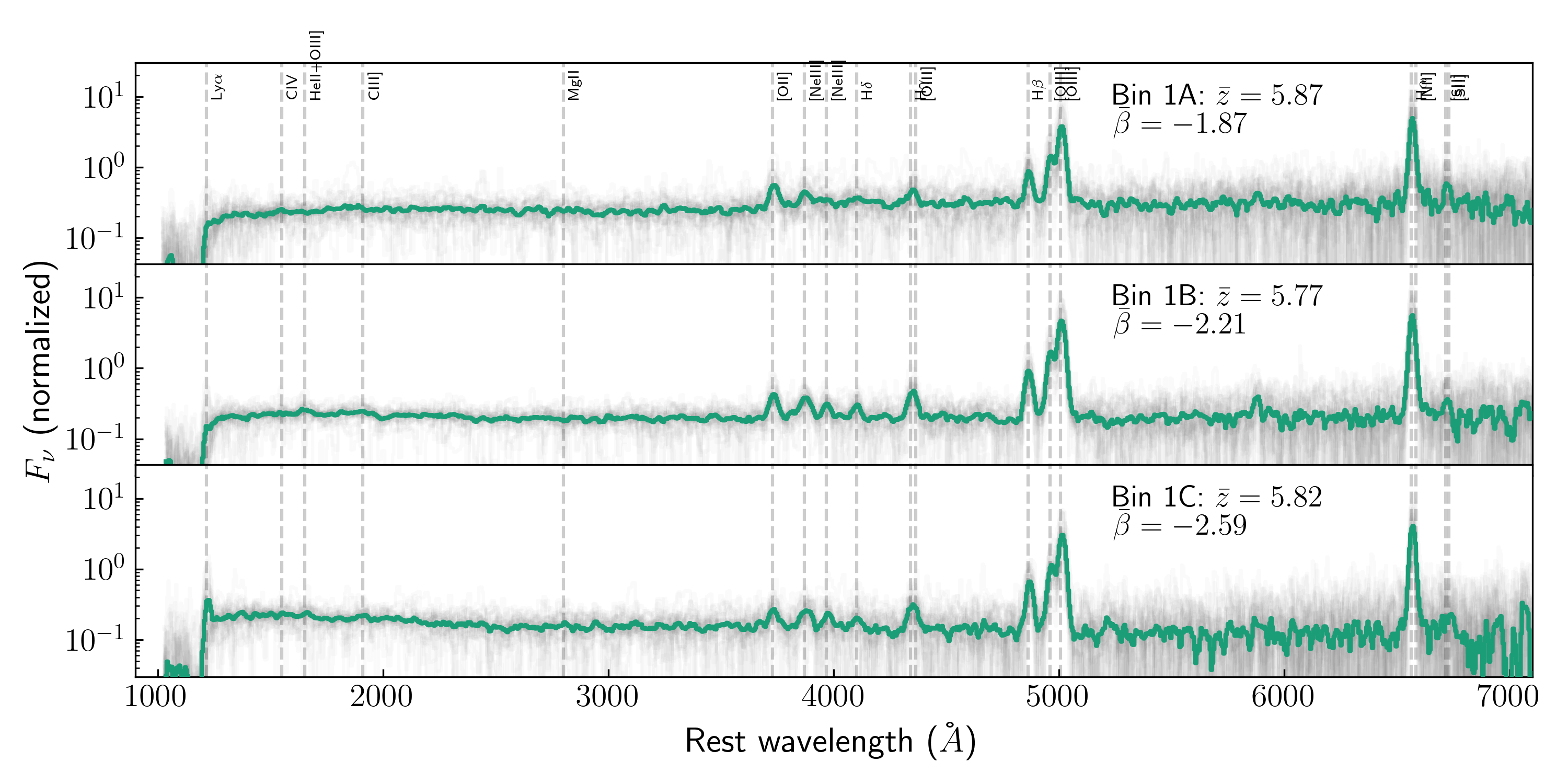}

    \includegraphics[width=\linewidth]{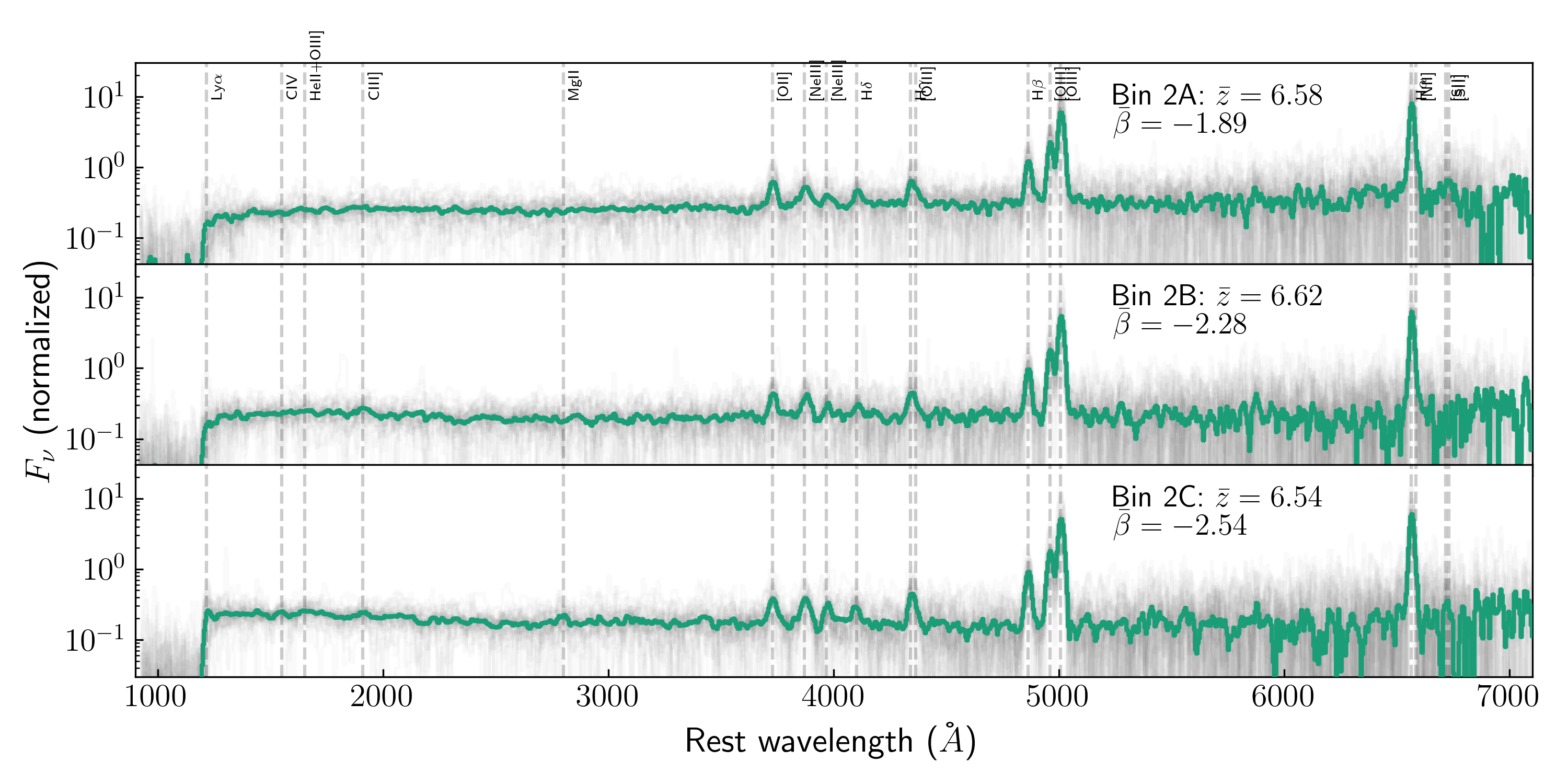}
    \caption{Stacked and normalized rest-frame spectra (in units of $F_\nu$) created in bins of redshift and $\beta$ (green), with individual spectra that go into the stack shown in gray in the background, as described in Section \ref{sec:data} and in Table \ref{tab:stack_bins}. The locations of key emission lines have been marked with dashed lines. Each bin shows the median redshift and the $\beta$ value measured from the stacked spectrum, which is consistent with the median $\beta$ in the bin.}    
    \label{fig:stacks1}
\end{figure*}

\begin{figure*}
    \centering
    \ContinuedFloat
    \includegraphics[width=\linewidth]{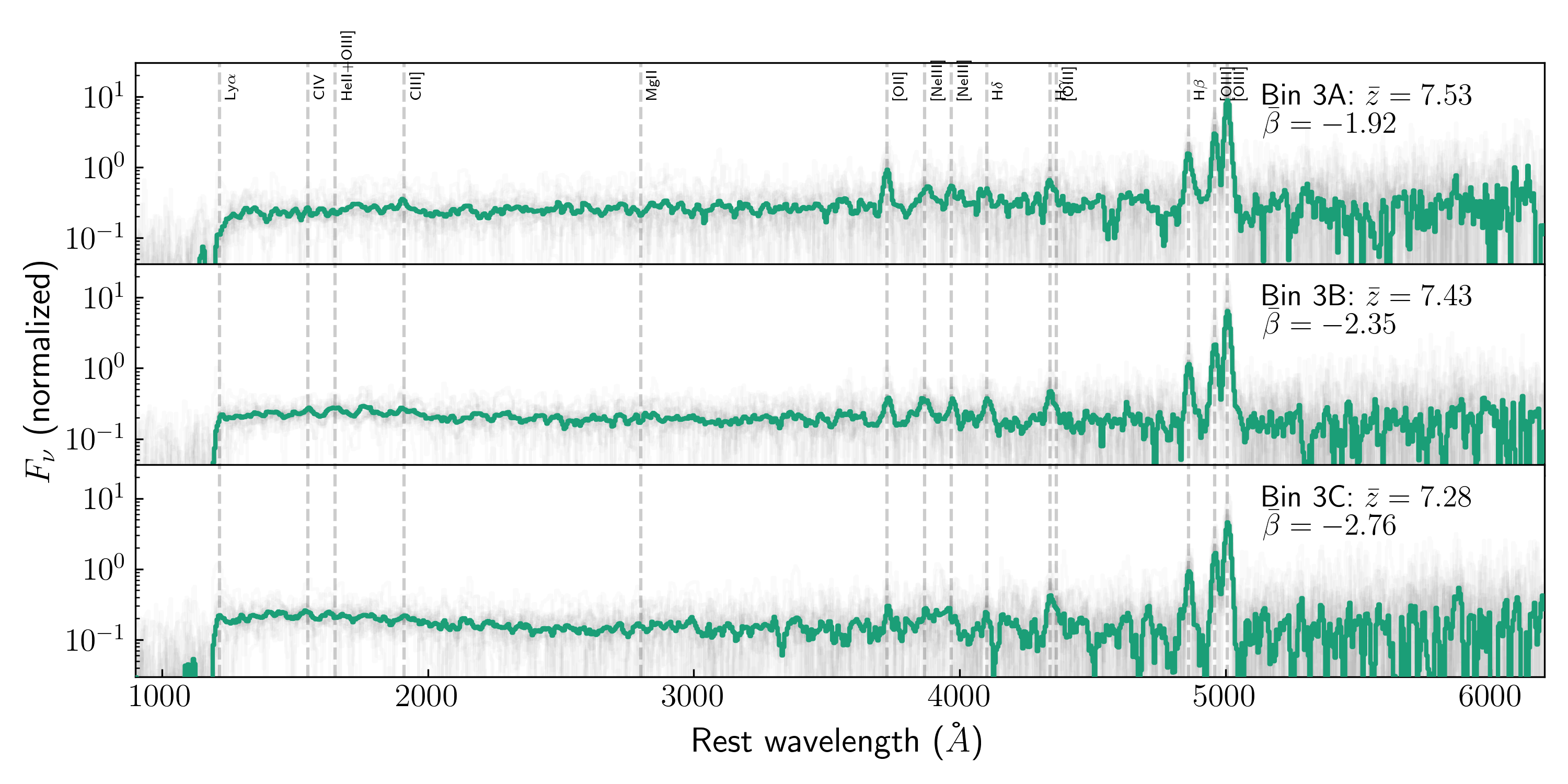}

    \includegraphics[width=\linewidth]{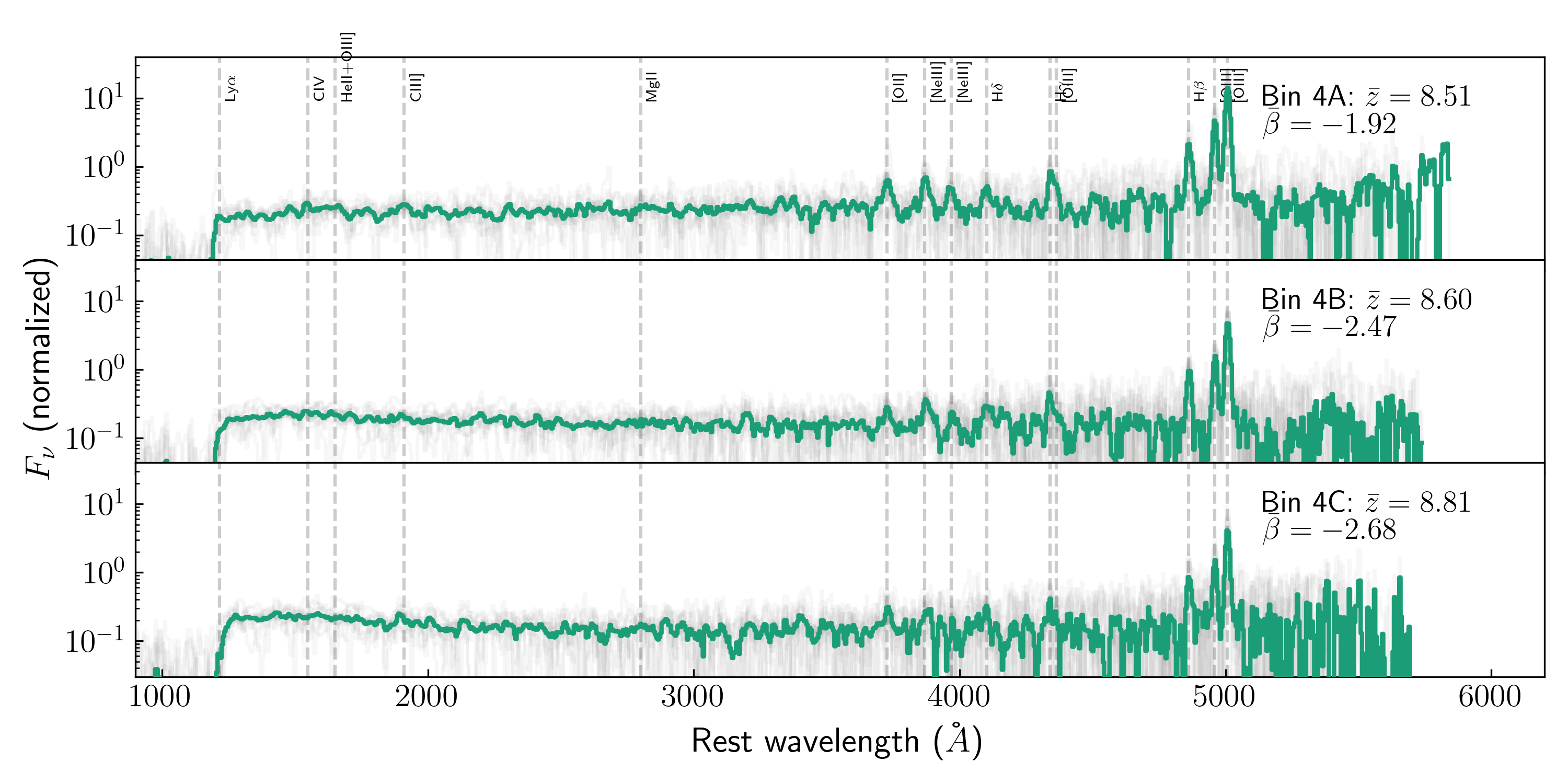}

    \includegraphics[width=\linewidth]{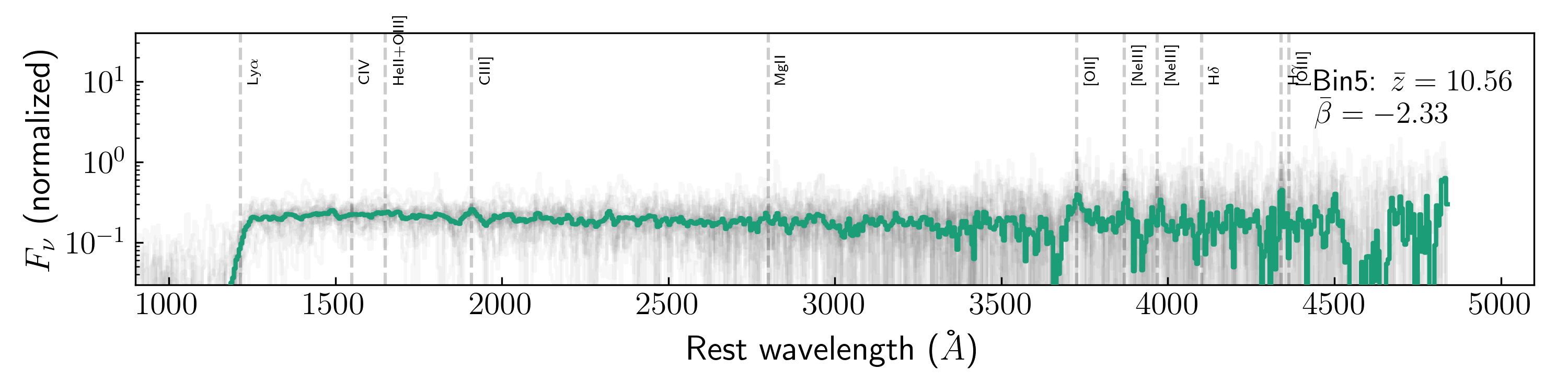}
    \caption{continued.}
\end{figure*}


\bsp	
\label{lastpage}
\end{document}